\documentclass[12pt, a4paper]{article}
\usepackage{amsmath}
\usepackage{graphicx,psfrag,epsf}
\usepackage{enumerate}
\usepackage[numbers]{natbib}

\usepackage[nodisplayskipstretch]{setspace}
\usepackage{amsmath}
\usepackage{array}
\usepackage{url}
\usepackage{graphicx,psfrag,epsf, amsmath}
\usepackage{enumerate}
\usepackage{booktabs}
\usepackage{tikz}

\newcommand{\blind}{0}
\usepackage{epstopdf,gensymb}
\usepackage{graphics}
\usepackage{graphicx,color}
\newcommand{\erdos}{Erd\H{o}s-R\'enyi }
\usepackage{amsfonts, animate,subfigure}
\usepackage[margin=1in]{geometry}
\usepackage{todonotes}
\usepackage{setspace}
\usepackage{indentfirst}
\usepackage{float}
\usepackage{amsfonts}
\usepackage{bm,pbox}
\usepackage{multirow}
\usepackage{color}
\usepackage{framed}
\usepackage{amssymb,amsthm,amsfonts,amsbsy,latexsym,dsfont}
\usepackage{enumerate}

 \newtheorem{theorem}{Theorem}
 
 \newtheorem{proposition}[theorem]{Proposition}

 \usepackage{etoolbox}

 \makeatletter
 \patchcmd{\@makecaption}
   {\parbox}
   {\advance\@tempdima-\fontdimen1} % decrease the width!
   {}{}
 \makeatother  

 \usepackage{graphicx}
 \usepackage[caption = false]{subfig}

\date{}

\begin{document}

\def\spacingset#1{\renewcommand{\baselinestretch}%
{#1}\small\normalsize} \spacingset{1}

\setlength{\abovedisplayskip}{3pt}
\setlength{\belowdisplayskip}{3pt}
%%%%%%%%%%%%%%%%%%%%%%%%%%%%%%%%%%%%%%%%%%%%%%%%%%%%%%%%%%%%%%%%%%%%%%%%%%%%%%

\if0\blind
{
  \title{\bf Varying-Coefficient Models for Dynamic Networks}
  \author{Jihui Lee\\
	Department of Biostatistics, Columbia University \\ 
%	New York, NY 10032 \url{jl4201@cumc.columbia.edu}
	Gen Li \thanks{
    The authors gratefully acknowledge that Gen Li's research was partially supported by the Calderone Junior Faculty Award from the Mailman School of Public Health at Columbia University.
    %\textit{please remember to list all relevant funding sources in the unblinded version}
    }\hspace{.2cm} \\
	Department of Biostatistics, Columbia University \\ 
%	New York, NY 10032 \url{gl2521@cumc.columbia.edu}
    and \\
    James D. Wilson \\
    Department of Mathematics and Statistics, University of San Francisco
%	San Francisco, CA 94117 \url{jdwilson4@usfca.edu}
}
  \maketitle
} \fi

\if1\blind
{
  \bigskip
  \bigskip
  \bigskip
  \begin{center}
    {\LARGE\bf Title}
\end{center}
  \medskip
} \fi

\bigskip
\begin{abstract}
Dynamic networks are commonly used in applications where relational data is observed over time. Statistical models for such data should capture not only the temporal dependencies between networks observed in time, but also the structural dependencies among the nodes and edges in each network. As a consequence, effectively making inference on dynamic networks is a computationally challenging task, and many models established for dynamic networks are intractable even for moderately sized networks. In this paper, we propose and investigate a family of dynamic network models, known as varying-coefficient exponential random graph models (VCERGMs), that characterize the evolution of network topology through smoothly varying parameters in an exponential family of distributions. The VCERGM provides an interpretable dynamic network model that enables the inference of temporal heterogeneity in a dynamic network. We establish how to fit the VCERGM through maximum pseudo-likelihood techniques, and thus provide a computationally tractable method for statistical inference of complex dynamic networks. We furthermore devise a bootstrap hypothesis testing framework for testing the temporal heterogeneity of an observed dynamic network sequence. We apply the VCERGM to the US Congress co-voting network and a resting-state brain connectivity case study and show that our method provides relevant and interpretable patterns describing each data set. Comprehensive simulation studies demonstrate the advantages of our proposed method over existing methods.
% The VCERGM builds on the family of exponential random graph models and provides an interpretable model for understanding the underlying heterogeneity of a dynamic network.

%Our model imposes smoothness on the varying parameters through a roughness penalty on the coefficients of a basis-spline expansion. 
% When modeling a dynamic network, accounting for its evolving structure is critical to making valid inferences and providing interpretable summaries of the network.
\end{abstract}

\noindent%

\spacingset{1.45}

%%%%%%%%%%%%%
% Section: Introduction  %
%%%%%%%%%%%%%

\newpage

\section{Introduction}\label{sec:intro}

\begin{figure}[htbp]
\centering
\subfigure[$40^{\text{th}}$ Congress]{\includegraphics[width = 0.3\textwidth]{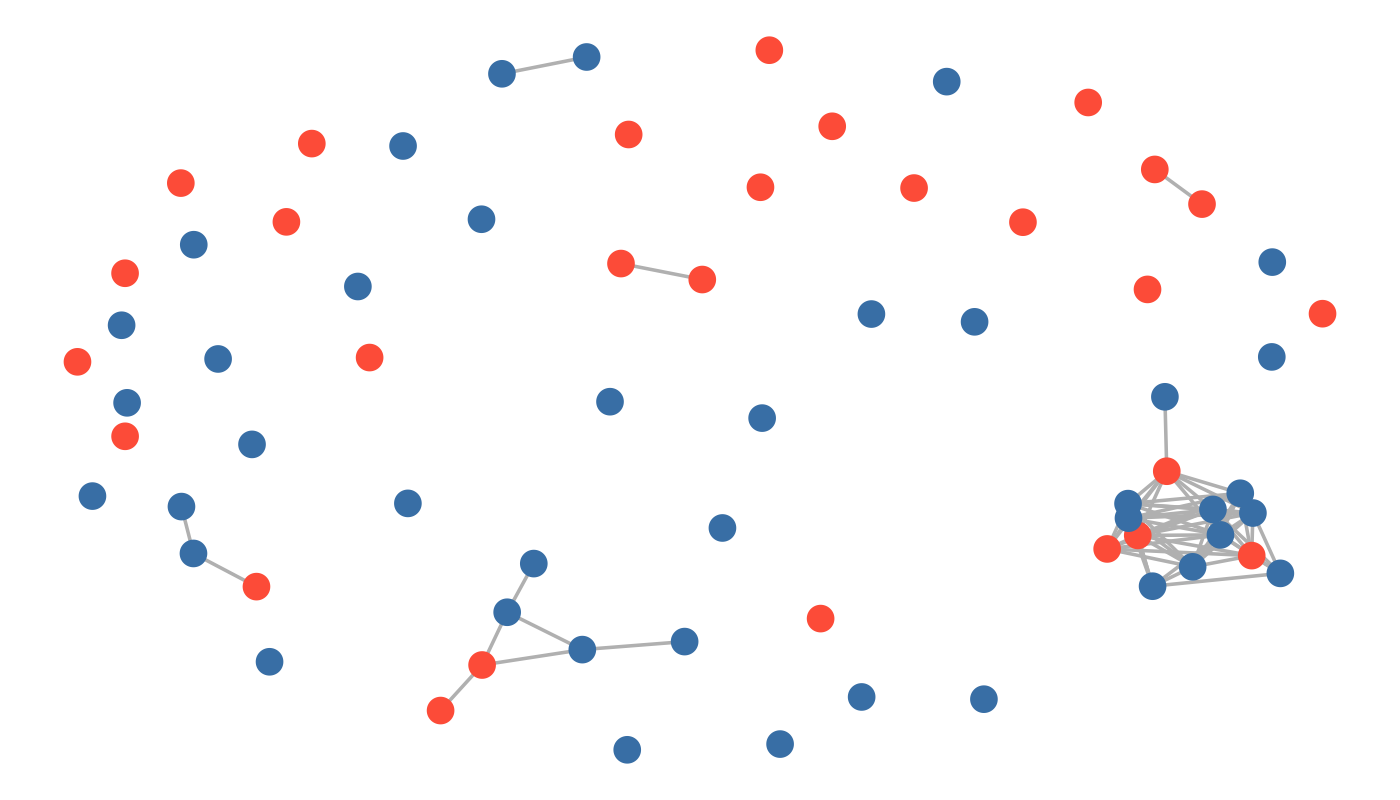}}
\subfigure[$70^{\text{th}}$ Congress]{\includegraphics[width = 0.3\textwidth]{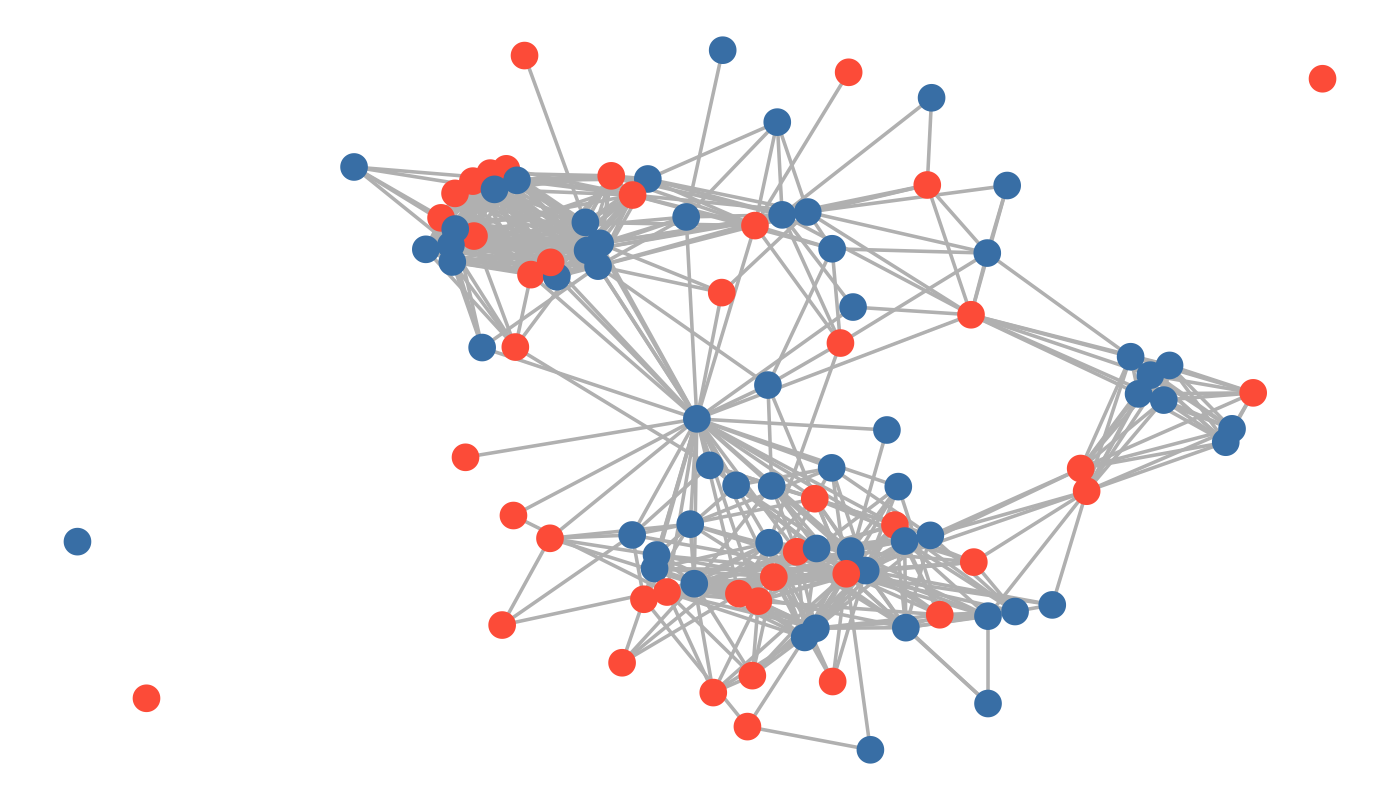}}
\subfigure[$113^{\text{th}}$ Congress]{\includegraphics[width = 0.3\textwidth]{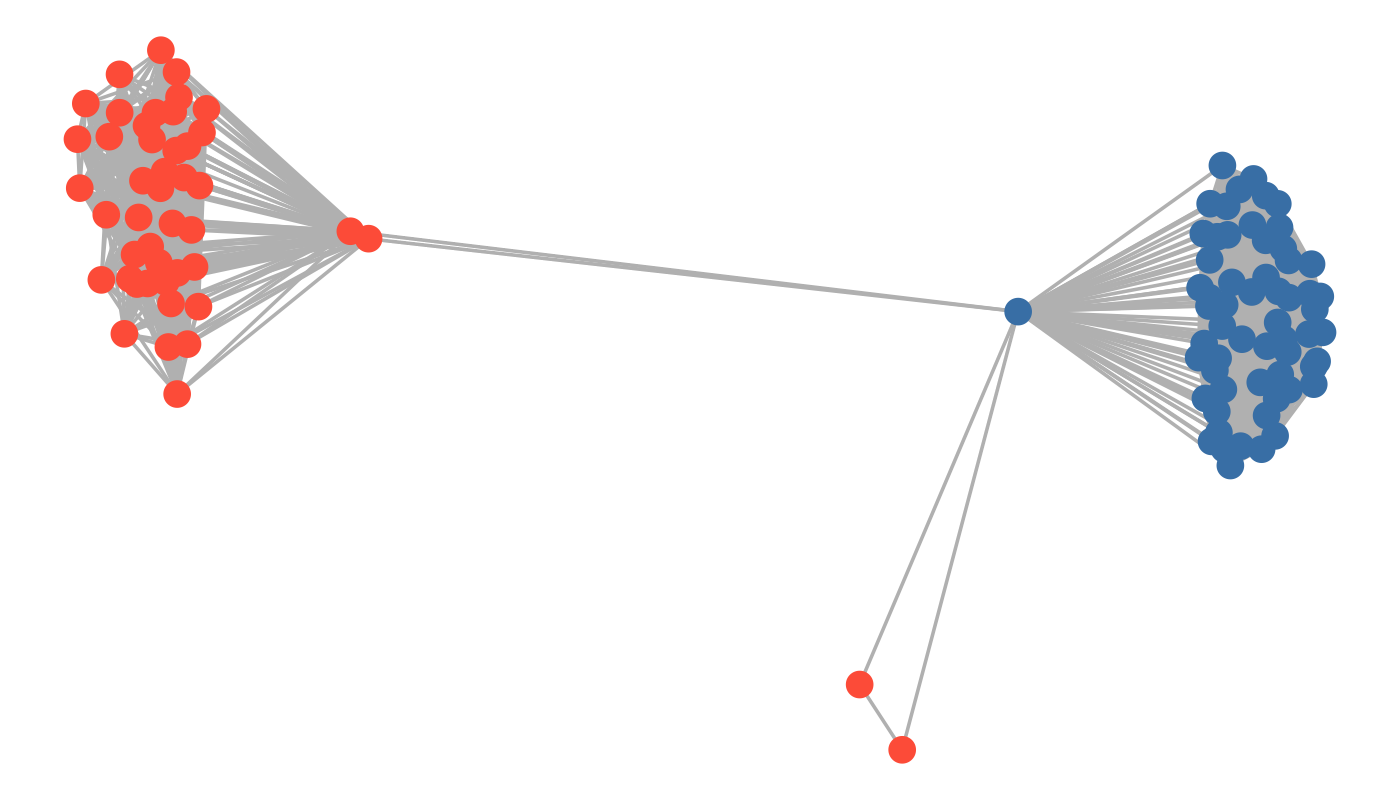}}\\
\caption{{\bf U.S. Senate co-voting network}: Co-voting networks of US senators in Congress 40, 70, and 113. Red nodes represent Republican Senators and blue nodes represent Democratic Senators.}
\label{figure:TemporalNetworks1}
\end{figure}

\begin{figure}[htbp]
\centering
\subfigure[Time 10]{\includegraphics[width = 0.3\textwidth]{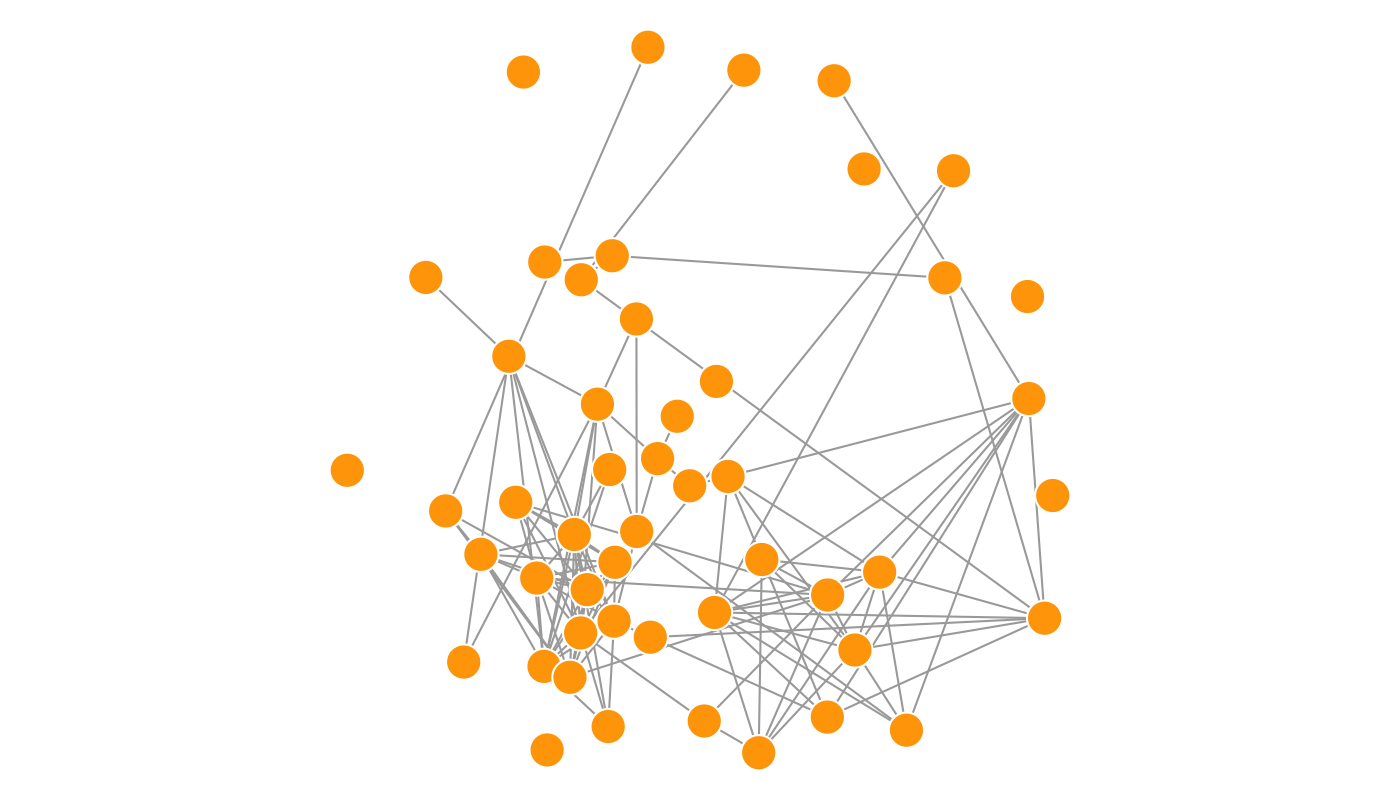}}
\subfigure[Time 20]{\includegraphics[width = 0.3\textwidth]{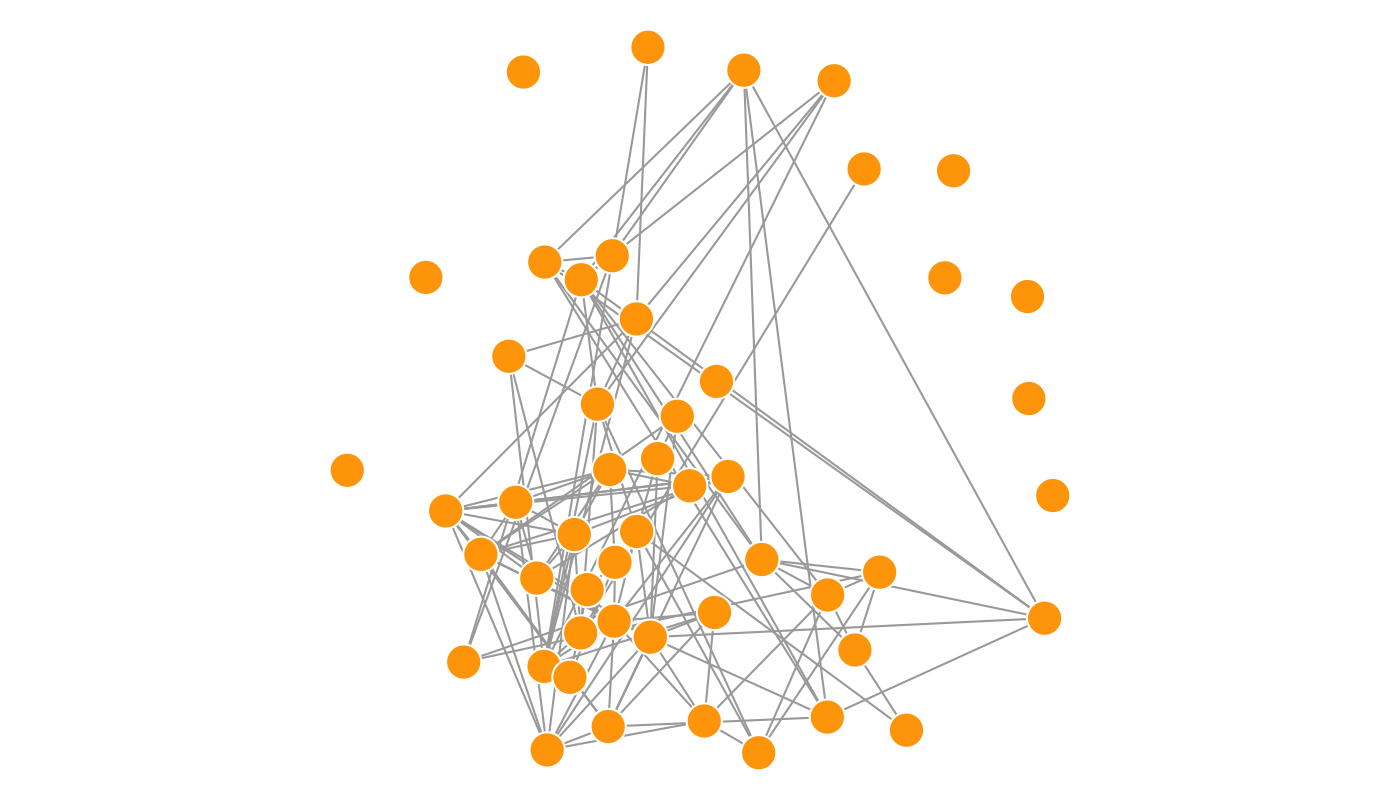}}
\subfigure[Time 47]{\includegraphics[width = 0.3\textwidth]{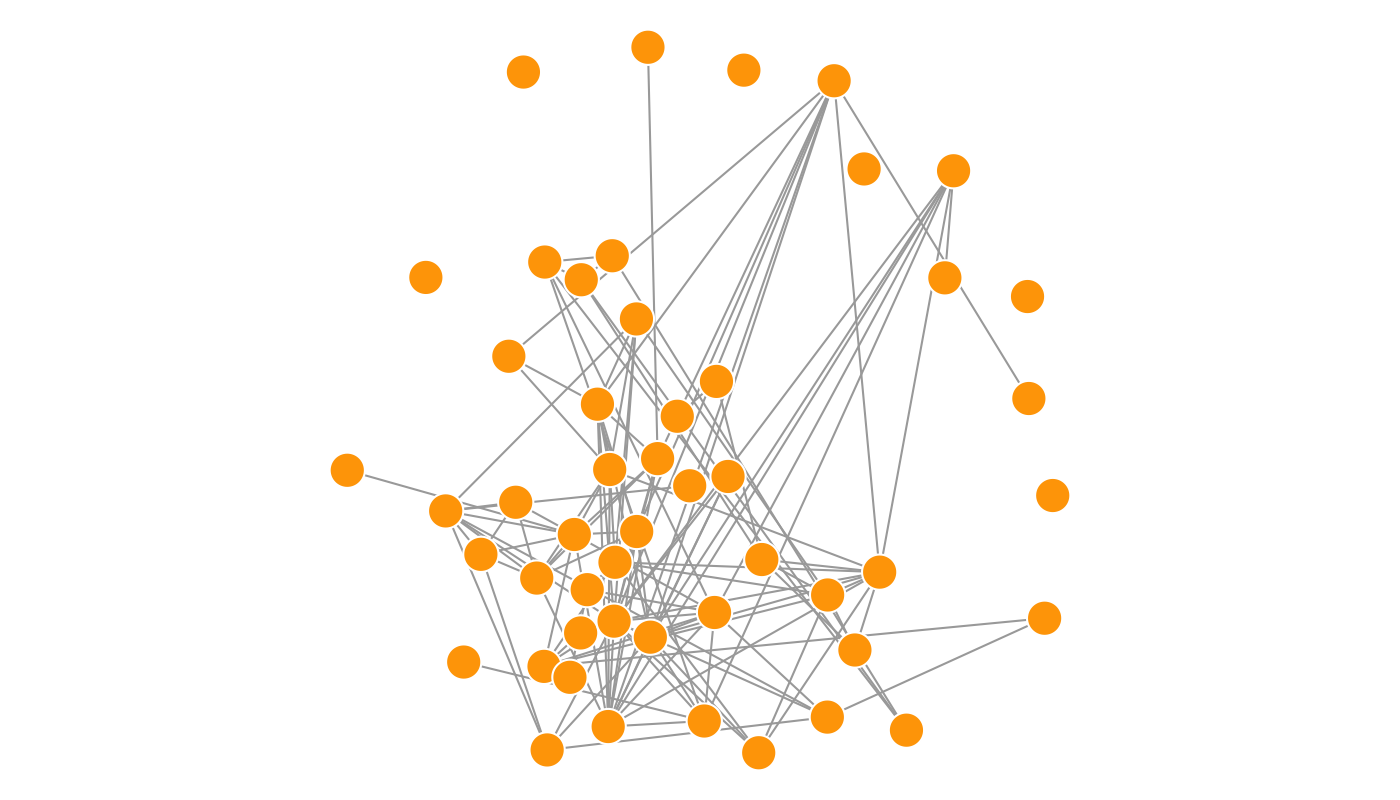}}
\caption{{\bf Resting state fMRI network}: Resting state fMRI network at observed times 10, 20, and 47. Each node represents a brain region. The top 10\% of partial precision between regions form an edge.}
\label{figure:TemporalNetworks2}
\end{figure}

% Importance of modeling temporal effects in networks
Networks have been extensively used to explore, model, and analyze the relational structure of individual units, or actors, in a complex system. In a network model, nodes represent the actors of the system, and edges are placed between nodes if the corresponding actors share a relationship. In many applications, the relationships among the actors of a modeled system change over time, necessitating the use of dynamic networks.
Two diverse examples, which we analyze later in our application study, include the Congressional co-voting networks in Figure \ref{figure:TemporalNetworks1} and resting state brain connectivity networks in Figure \ref{figure:TemporalNetworks2}. 
%to the analysis of functional brain activity of a person in resting state (Figure \ref{figure:TemporalNetworks2}) \citep{cabral2014exploring, moody2013portrait}.
A prominent way to analyze relational network systems is through the use of probabilistic models, or graphical models, which describe the generative mechanism of an observed network. Although there is a rich body of literature on graphical models for static networks \citep{fienberg2012brief, goldenberg2010survey}, the development of interpretable and computationally tractable models for dynamic networks is in its early stages.

% Figure 1: (Example of dynamic networks) Political networks

% Figure 2: (Example of dynamic networks) fMRI networks

% Temporal heterogeneity (The importance of modeling heterogeneity)
An important feature of dynamic networks that needs to be captured in any statistical model is the extent to which its local and global features change through time. We refer to this property as \emph{temporal heterogeneity}. Heterogeneity directly affects the underlying process that best describes the formation of network. In parametric models, heterogeneity may result in significant changes in parameters that characterize the observed network. Consider the U.S. Senate co-voting network shown in Figure \ref{figure:TemporalNetworks1}. One can readily observe an evolution of the network to form distinct clusters of Republicans and Democrats by the 113th Congress. Moreover, this configuration is in stark contrast with the sparse, seemingly random configuration formed in the 40th Congress. On the other hand, the resting state  functional magnetic resonance imaging (fMRI) network shown in Figure \ref{figure:TemporalNetworks2} remains fairly stable through time with only minor local changes in edge formation. These contrasting examples exemplify the need to explicitly model the heterogeneity of a network. We further analyze these dynamic networks in Sections \ref{subsec:political} and \ref{subsec:fMRI}.

% Introduce our proposed method
The examples above highlight the importance of accounting for evolving structure when modeling a dynamic network. In this paper, we propose a probabilistic model for dynamic networks called the varying-coefficient exponential random graph model (VCERGM). The model parameterizes time-varying topological features of dynamic networks in continuous time. Our model builds on two major statistical methodologies. One is the exponential family of random graph models \citep{holland1981exponential, wasserman:1996} that characterizes the marginal effect of local and global network features on the likelihood of the network. The other major component is a varying-coefficient specification \citep{hastie1993varying}, which flexibly models the changes of effect parameters over time. The VCERGM characterizes the temporal heterogeneity of dynamic network by modeling the parameter associated with each topological feature as a smooth function of time.

% Advantages / benefits of VCERGM
By quantifying temporal heterogeneity of a network via fluctuating parameters, we are able to analyze key properties of the local and global features of a dynamic network through the VCERGM. In addition to serving as a means to test for heterogeneity of a dynamic sequence, our model can also be directly used for interpolation of missing networks or edges. For networks at unobserved time points, the VCERGM provides robust estimates that reflect the structure of the unobserved networks without being strongly influenced by outliers in the sequence. Importantly, estimation of the VCERGM can be done with maximum pseudo-likelihood estimation (MPLE), enabling computationally efficient inference for dynamic networks with many nodes or time points.

% VCERGM as an extension of ERGM and TERGM
Several dynamic network models that have been investigated. We briefly describe these here.
The exponential random graph model (ERGM) is a family of probability distributions on unweighted static network. The ERGM has been adapted to dynamic networks in the pivotal work of \cite{hanneke2010discrete}. The method is called the temporal exponential random graph model (TERGM). The TERGM models the difference in topological features between every two consecutive networks in a similar fashion to the ERGM. However, it ignores the heterogeneity of the differences, and cannot fully capture the time-varying patterns of the network structure. In fact, we show that in a wide range of situations the TERGM degenerates to a collection of independent and identically distributed ERGMs (see supplementary materials). 

The TERGM has been further investigated in many different perspectives. \cite{guo2007recovering} devised the hidden TERGM (HTERGM), which utilizes a hidden Markov process to express the nature of rewiring networks and model a time-specific network topology. \cite{krivitsky2014separable} generalized the TERGM to the separable TERGM (STERGM). The STERGM flexibly models the formation and dissolution of networks by separately parameterizing prevalence and duration of fluctuations. 

% However, like the TERGM, the STERGM is essentially a special case of first-order TERGM that it still cannot capture the temporal heterogeneity.

Another method for dynamic network modeling is the stochastic actor-oriented model (SAOM) \citep{snijders2001statistical}. It provides an alternative to dyadic models and instead is a localized actor-based model, which characterizes network evolution as a consequence of each actors' connectivity. Even if the SAOM considers the fluctuation between two time points, it does not provide explicit form to parametrize the fluctuation in network topology. \cite{sarkar2005dynamic} and \cite{sewell2015latent} generalized the latent space model developed by \cite{hoff2002latent} to dynamic networks. The dynamics of network structure is modeled through random effects in a latent space. It focuses on the transition between two time points and provides limited description on overall network.

% Time-varying models
Compared to time-invariant models described above, an alternative actor-based model was introduced in \cite{hoff2015multilinear}, where dynamic networks are modeled using multilinear tensor regression. This work adapted autoregressive models to dictate temporal dependence in a sequence of networks, and like the SAOM, proposed an actor-based dependence structure between edges in each network. It directly models the temporal heterogeneity but may not be adequate for larger networks due to its computational complexity. In the meantime, \cite{kolar2010estimating} emphasizes on capturing time-varying attributes of dynamic networks and parametrizes the evolving relationship of each edge between nodes as a smooth function of time. Along with kernel smoothing approach, the $\ell_1$-regularization is utilized to ensure the smoothness. The parameters in the model provide a valuable intuition in understanding the topological change of each edge, but fitting this model for larger networks can be computationally expensive considering the number of parameters.

%%%I AM HERE%%%
The VCERGM exploits a varying-coefficient framework to model the temporal heterogeneity of topological features.
In general, varying-coefficient models form a family of semi-parametric models in which the coefficient of some parametric model evolves with characteristics of the coefficient in a nonparametric manner.
Varying-coefficient models were first used to characterize non-linear effects of covariates on real-valued response variables \citep{hastie1993varying}.
Later it was extended to the dynamic generalized linear models \citep{hoover1998nonparametric, zhang2014varying}. A detailed review of varying-coefficient models and their applications are provided in \cite{fan2008statistical}.
In our proposed model, we model the coefficients of the topological features of a network as a function of time. The marginal parametric model takes the form of an exponential random graph model (ERGM). In this setting, the varying coefficients effectively capture the dynamic pattern of network structure by modeling the variation in the sufficient statistics of an observed sequence of networks. 
% To our best knowledge, the VCERGM is the first attempt to generalize the idea to modeling dynamic networks.

The remainder of this paper is organized as follows. In Section \ref{sec:model} we introduce the varying-coefficient exponential random graph model for dynamic networks. Section \ref{sec:estimation} describes the pseudo-likelihood estimation of the VCERGM via penalized logistic regression. A hypothesis testing procedure for formally testing temporal fluctuation in a sequence of time-varying graphs is introduced in Section \ref{sec:hypothesis}. We assess the performance of the VCERGM on a series of simulated dynamic networks, comparing it to alternative generative graph models in Section \ref{sec:simulations}. In Section \ref{sec:applications}, we apply the VCERGM to the functional magnetic resonance imaging (fMRI) and co-voting networks and analyze the dynamic nature of each of these applications.
%we further investigate the utility of the VCERGM through two case studies: a functional magnetic resonance imaging (fMRI) study on 500 patients from the WU-Minn Consortium Human Connectome Project, and the co-voting behavior of U.S. Senators from 1857 to 2013.
We conclude with a discussion of our method and open areas for future research in Section \ref{sec:discussion}.

%%%%%%%%%%
% Section: Model %
%%%%%%%%%%

\section{Model}\label{sec:model}
We begin by describing the exponential family of random graph models (ERGMs) and their temporal extension, the TERGM, since our proposed model is closely related to these specifications. We then describe our proposed model, the VCERGM.
\subsection{Exponential Random Graph Models}
% ERGM: Model specification
Suppose that $X \in \{0,1\}^{n \times n}$ is an unweighted network with $n$ vertices, whose $(i,j)$th entry $X_{ij}$ is an indicator that specifies whether or not node $i$ and node $j$ are connected by an edge. The ERGM is a probability distribution that characterizes the likelihood of $X$ via a function of network statistics $\mathbf{h}: \{0,1\}^{n \times n} \rightarrow \mathbb{R}^p$ that describe the topological structure of $X$. One may include, for example, a statistic that quantifies the tendency of the nodes in the network to exhibit reciprocal ties using the reciprocity statistic $h_1(X) = \sum_{i \neq j} X_{ij} X_{ji}$. \cite{hunter2008goodness} describes a means of determining which network statistics to include in a model based on collection of goodness of fit diagnostics. Given $\mathbf{h}$, the ERGM posits that $X$ is a binary random matrix generated from the following probability mass function
\begin{equation}\label{eq:ERGM}
  \mathbb{P}(X = x \mid \boldsymbol{\phi}) = \frac{\exp\{\boldsymbol{\phi}^T \mathbf{h}(x)\}}{\displaystyle\sum_{z \in \{0,1\}^{n \times n}} \exp\{\boldsymbol{\phi}^T \mathbf{h}(z)\}},
\end{equation}
\noindent where $\boldsymbol{\phi} \in \mathbb{R}^p$ parameterizes the influence of the network statistics $\mathbf{h}(X)$ on the likelihood of $X$. The ERGM has been successfully applied in a wide variety of fields, ranging from social networks to brain connectivity networks \citep{goodreau2009birds, simpson2011exponential, szekely2016childhood}. Recent tutorials of exponential random graph models and their applications are provided in \cite{cranmer2011a, fellows2012exponential, robins2007introduction}.

% TERGM: Model specification
The TERGM is an important and popular statistical model for inference of dynamic networks \citep{desmarais2012statistical, hanneke2010discrete}, and can be described as follows. Consider a dynamic network $\mathbf{X} = \{X_1, X_2, \ldots, X_T\}$ that is observed at $T$ discrete and non-overlapping time periods, where each graph $X_t \in \{0,1\}^{n \times n}$ from $\mathbf{X}$ is unweighted, and observed for the set of vertices $[n] = \{1, \ldots, n\}$. The TERGM is a generative model for $\mathbf{X}$ that characterizes the conditional probability of $X_t$ given $\mathbf{X}_t^{-} = \{X_s: s = 1, \ldots, t-1\}$ via an exponential family of probability distributions. Under the first order TERGM, $\mathbf{X}$ exhibits a one-step Markov dependence between sequential networks as follows.
\begin{equation}\label{eq:MarkovAssumption}
	\mathbb{P}(X_t = x_t \mid {\bf X^{-}_{t}} = {\bf x^{-}_t}) = \mathbb{P}(X_t = x_t \mid X_{t-1} = x_{t-1}) %\tag{A1}
\end{equation}
Under (\ref{eq:MarkovAssumption}), one can fully specify the joint probability mass function of $\mathbf{X}$ by parameterizing the one-step transitions from $X_{t-1}$ to $X_t$. One models these dependencies using a function of transition statistics $\mathbf{g}: \{0,1\}^{n\times n} \times \{0,1\}^{n\times n} \rightarrow \mathbb{R}^p$. These statistics represent the temporal potential over cliques across two sequential networks and can represent, for example, the change in the clustering or the change in overall connectivity between each pair of networks. For a chosen $\mathbf{g}$, the first-order TERGM specifies the likelihood of $X_t \mid X_{t-1}$ for $t = 2, \ldots, T$ as
\begin{equation}\label{eq:tergm}
		\mathbb{P}(X_t = x_t \mid \mathbf{X}_t^{-}; \boldsymbol{\phi}) = \mathbb{P}(X_t = x_t \mid X_{t-1} = x_{t-1}; \boldsymbol{\phi}) = \dfrac{\exp\{\boldsymbol{\phi}^T ~ \mathbf{g}(x_t, x_{t-1})\}}{\displaystyle\sum_{z \in \{0,1\}^{n \times n}} \exp\{\boldsymbol{\phi}^T ~ \mathbf{g}(z, x_{t-1})\}},
\end{equation}
\noindent where $\boldsymbol{\phi} \in \mathbb{R}^p$ parameterizes the influence of the transition statistics $\mathbf{g}(X_{t}, X_{t-1})$ on the conditional likelihood of $X_t$ given $X_{t-1}$. Suppose that the marginal distribution $\mathbb{P}(X_1 = x_1 \mid \boldsymbol{\phi})$ is specified. The TERGM characterizes the joint distribution of the dynamic sequence $\mathbf{X}$ by
\begin{equation}\label{eq:tergm2}
	\mathbb{P}(\mathbf{X} = \mathbf{x} \mid \boldsymbol{\phi}) = \mathbb{P}(X_1 = x_1 \mid \boldsymbol{\phi}) \prod_{t = 2}^T \mathbb{P}(X_t = x_t \mid X_{t-1} = x_{t-1}, \boldsymbol{\phi}).
\end{equation}
We note that in general if one is able to specify appropriate transition statistics, then the TERGM in (\ref{eq:tergm}) and (\ref{eq:tergm2}) is readily generalized to higher-order Markov dependency; however, for discussion of our equivalence statement in the supplementary materials, we consider the first-order TERGM only.

\subsection{Varying-Coefficient Exponential Random Graph Models}

% What is different in VCERGM compared to other existing methods
Let $\mathbf{X} = \{X_t: 0 \leq t \leq T\}$  be a stochastic sequence of temporally ordered networks observed continuously up to some time $T > 0$. At each time point $t$, $X_t\in\{0,1\}^{n_t\times n_t}$ represents an unweighted, directed or undirected network with $n_t$ nodes. Our goal is to provide a dynamic network model for $\mathbf{X}$ that directly accounts for the temporal heterogeneity of its local and global network structure.
%To do so, we extend the notion of varying-coefficient models from \cite{hastie1993varying} to the family of exponential random graph models.

The VCERGM consists of two components - (i) an ERGM representation for the marginal likelihood of each observed network, and (ii) the coupling of networks over time via a varying-coefficient model, where the coefficients at time $t$ parameterize the marginal likelihood of the network $X_t$. We first specify a set of functions $\mathbf{h}(x_t, n_t): \{0,1\}^{n_t \times n_t} \rightarrow \mathbb{R}^p$ for $t\in[0,T]$, which quantify the same $p$ topological features of network $x_t$ with size $n_t$. Given $\mathbf{h}(x_t, n_t)$ and the coefficient vector $\boldsymbol{\phi}(t)=(\phi_1(t),\cdots,\phi_p(t))^T\in\mathbb{R}^p$, the marginal likelihood of $X_t$ at time $t$ has an ERGM representation given by
\begin{equation} \label{eq:VCERGM}
\mathbb{P}(X_t = x_t \mid \boldsymbol{\phi}(t)) = \frac{\exp\{\boldsymbol{\phi}(t)^T\mathbf{h}(x_t, n_t)\}}{\sum_{z \in \{ 0,  1\} ^{n_t \times n_t}} \exp \{ \boldsymbol{\phi}(t)^T \, \mathbf{h}(z, n_t) \}}, \text{\hskip 1pc $x_t \in \{0,1\}^{n_t \times n_t}$}.\\
% \theta_j(t) &= \sum_{\ell = 1}^q \Phi_{j\ell} B_{\ell}(t), \text{\hskip 1pc $j = 1, \ldots, p$}\\
\end{equation}
A large collection of topological features can be used in the VCERGM. Traditionally, the network statistics are raw counts of different features in an observed network, such as the number of edges (edge density), the number of triangles, or the number of reciprocal edges in a directed network.
%For example, the number of triangles in an undirected graph provides a measure of clustering in that graph. On the other hand, the number of bi-directional relationships in a directed graph provides a measure of reciprocity in the graph.
However, in dynamic networks, networks at different time points may have differing numbers of nodes. Thus it is not appropriate to use the raw counts directly.
Instead, one should standardize the network statistics to make them comparable over time.
We propose to standardize the raw count of each feature by its maximal possible value, and use the standardized statistics in model \eqref{eq:VCERGM}.
Table \ref{table:NetworkStatistics} provides examples of standardized network statistics for a binary graph $X_t$ with $n_t$ nodes, where $x^t_{ij}$ represents whether there is an edge from node $i$ to node $j$ in $X_t$.
%For illustration, we provide three relevant statistics for directed networks - edge density, reciprocity, and cyclic triads - in.

\begin{table}[htbp]
\caption{\label{table:NetworkStatistics} Examples of network statistics that can be specified in the VCERGM}
\centering
\resizebox{.7\columnwidth}{!}{%
%\fbox{%
\begin{tabular}{clcl}
\toprule
Type & \multicolumn{2}{c}{Network Statistic} & \multicolumn{1}{c}{Definition} \\
\hline
Directed & Edge density &
\begin{tikzpicture}[shorten >=1pt,->]
  \tikzstyle{vertex}=[circle,fill=black!25,minimum size=17pt,inner sep=0pt]
  \foreach \name/\x in {i/1, j/2}
    \node[vertex] (G-\name) at (\x,0) {$\name$};
  \foreach \from/\to in {i/j} \draw (G-\from) -- (G-\to);
\end{tikzpicture} &
$\displaystyle \sum_{i \neq j} x^t_{ij} \, / \, \{n_t (n_t - 1) \}$ \\
& Reciprocity &
\begin{tikzpicture}[shorten >=1pt,->]
  \tikzstyle{vertex}=[circle,fill=black!25,minimum size=17pt,inner sep=0pt]
  \foreach \name/\x in {i/1, j/2}
    \node[vertex] (G-\name) at (\x,0) {$\name$};
  \foreach \from/\to in {i/j} \draw (G-\from) -- (G-\to);
    \foreach \from/\to in {j/i} \draw (G-\from) -- (G-\to);
\end{tikzpicture} &
$\displaystyle \left. \sum_{i < j} x^t_{ij} \, x^t_{ji} \, \middle/ \, \binom{n_t}{2} \right. $ \\
& Cyclic triad &
\begin{tikzpicture}[scale = 0.7, shorten >=1pt,->]
  \tikzstyle{vertex}=[circle,fill=black!25,minimum size=17pt,inner sep=0pt]
  \foreach \name/\angle/\text in {G-1/180/i, G-2/90/j, G-3/0/k}
    \node[vertex,xshift=6cm,yshift=.5cm] (\name) at (\angle:1cm) {$\text$};
  \foreach \from/\to in {1/2, 2/3, 3/1}
    \draw (G-\from) -- (G-\to);
\end{tikzpicture} &
$\displaystyle  \left. \sum_{i < j < k} x^t_{ij} \, x^t_{jk} \, x^t_{ki} \, \middle/ \, \left\{ \binom{n_t}{3} \times 2 \right\} \right.$ \\
%\hline
Undirected  & Two-star &
\begin{tikzpicture}[scale = 0.7, shorten >=1pt,-]
  \tikzstyle{vertex}=[circle,fill=black!25,minimum size=17pt,inner sep=0pt]
  \foreach \name/\angle/\text in {G-1/180/i, G-2/90/j, G-3/0/k}
    \node[vertex,xshift=6cm,yshift=.5cm] (\name) at (\angle:1cm) {$\text$};
  \foreach \from/\to in {1/2, 2/3}
    \draw (G-\from) -- (G-\to);
\end{tikzpicture} &
$\displaystyle  \left. \sum_{i < j < k} x^t_{ij} \, x^t_{jk} \, \middle/ \, \left\{ \binom{n_t}{3} \times 3 \right \} \right.$ \\
& Triangle &
\begin{tikzpicture}[scale = 0.7, shorten >=1pt,-]
  \tikzstyle{vertex}=[circle,fill=black!25,minimum size=17pt,inner sep=0pt]
  \foreach \name/\angle/\text in {G-1/180/i, G-2/90/j, G-3/0/k}
    \node[vertex,xshift=6cm,yshift=.5cm] (\name) at (\angle:1cm) {$\text$};
  \foreach \from/\to in {1/2, 2/3, 3/1}
    \draw (G-\from) -- (G-\to);
\end{tikzpicture} &
$\displaystyle  \left. \sum_{i < j < k} x^t_{ij} \, x^t_{jk} \, x^t_{ki} \, \middle/ \, \binom{n_t}{3} \right.$ \\

% In-Two-Stars & $\displaystyle \sum_{i} \sum_{j < k, \, j \neq i, \, k \neq i} x^s_{ji} \, x^s_{ki}$ \\
% Out-Two-Stars & $\displaystyle \sum_{i} \sum_{j < k, \, j \neq i, \, k \neq i} x^s_{ij} \, x^s_{ik}$ \\
\bottomrule
\end{tabular}}
%}
\end{table}

The coefficients $\boldsymbol{\phi}(t)$ in model \eqref{eq:VCERGM} characterize the influence of the corresponding network statistics on determining the network structure.
When a dynamic network evolves gradually over time, it is reasonable to believe the coefficients will also change gradually. In such a case, $\boldsymbol{\phi}(t)$ can be represented by smooth functions of $t$ with continuous second order derivatives over $[0,T]$ \citep{ramsay2006functional}.
The temporal dependence among the graphs in $\mathbf{X}$ are captured by the dynamic pattern of the coefficient functions.
This is the fundamental assumption of our model.
In the special case where all the separate functions in $\boldsymbol{\phi}(t)$ are constant, the generative models underlying the dynamic networks are identical over time and
the VCERGM reduces to a family of marginally identically distributed ERGMs.
In Section \ref{sec:hypothesis}, we introduce a formal hypothesis testing procedure to test the temporal heterogeneity of the coefficients.

%Interpretation of these coefficients is straightforward. Consider a concrete example where we specify only an edge statistic, $h_s(x_s) = \sum_{i,j \in [n_s]} x_{ij}(s)$ and the corresponding coefficient is a linear function $\phi(t) = \alpha t$. Let $s' > s$ and suppose that the difference between the number of edges in each graph is $h_{s'}(x_{s'}) - h_s(x_s) = \delta$. In this case, the odds of observing the graph $x_{s'}$ with respect to observing the graph $x_s$ is $\exp\{\alpha \delta (s' - s)\}$. Thus when $\alpha > 0$, the probability of observing a graph with more edges at time $s'$ than at time $s$ is increased; whereas, the probability of observing a graph with fewer edges at time $s'$ than at time $s$ is decreased. On the other hand, when $\alpha < 0$, graphs at time $s'$ with fewer edges than at time $s$ are favored.
%{\color{red} Before introducing the basis spline, motivate why we would like to model theta(t) in a nonparametric way, and mention different possibilities.}
%\cite{krivitsky2014separable}
%In general, the functional representation of $\boldsymbol{\phi}(t)$ is flexible.

%%%%%%%%%%%%%
% Section: Estimation   %
%%%%%%%%%%%%%

\section{Estimation}\label{sec:estimation}

\subsection{Spline-Based Representation of Time-Varying Coefficients}\label{subsec:spline}

Without any constraint, the collection of coefficients $\{ \boldsymbol{\phi}(t): 0 \leq t \leq T \}$ contain an infinite number of parameters, making inference on (\ref{eq:VCERGM}) intractable.
To address this problem, we approximately represent these smooth functions as a linear combination of basis functions. Possible strategies of defining basis functions include piecewise polynomials \citep{de1978practical}, Fourier series \citep{konidaris2008value}, and wavelets \citep{daubechies1992ten}. For inferential purposes, we employ basis splines (b-splines) \citep{de1978practical, eilers1996flexible} as a way to reduce the dimensionality of estimation. B-splines are commonly used due to its flexibility in incorporating smoothing constraints.
% One can efficiently represent these smooth functions through the use of a linear combination of basis functions, a representation commonly known as basis splines (b-splines) .

In particular, we first specify a collection of basis functions $B_1(t), \ldots, B_q(t)$, $0 \leq t \leq T$, and then approximate $\phi_k(t)$ by a linear combination of these functions
$$\phi_k(t) = \sum_{\ell = 1}^q \Phi_{k \ell} \, B_{\ell}(t),$$
where $\Phi_{k\ell}$ quantifies the contribution of the $\ell$th basis function on $\phi_k(t)$. Let $\boldsymbol{\Phi}=(\Phi_{k\ell}; \, k=1, \ldots, p, \, \ell=1, \ldots, q)$ denote the $p \times q$ basis coefficient matrix and let $\mathbf{B}(t) = (B_1(t), \ldots, B_q(t))^T$ be the length $q$ vector of basis functions. We can represent the coefficients $\boldsymbol{\phi}(t)$ as
\begin{equation}\label{eq:basis_representation}\boldsymbol{\phi}(t) = \boldsymbol{\Phi}\mathbf{B}(t).\end{equation}
\noindent The set of $q$ basis functions represents the smoothness of $\boldsymbol{\phi}(t)$, and the coefficient matrix $\boldsymbol{\Phi}$ determines the shape and trajectory of the fluctuations through time. Under the basis representation in (\ref{eq:basis_representation}), the distribution of $\mathbf{X}_t$ in (\ref{eq:VCERGM}) is fully specified by the $pq$ parameters in the coefficient matrix $\boldsymbol{\Phi}$.

\subsection{Fast Estimation via Maximum Pseudo Likelihood}\label{subsec:MPLE}

% Introduce MPLE
For an observed dynamic sequence of unweighted graphs $\mathbf{x} = \{ x_s\in\{0,1\}^{n_s\times n_s}: s = t_1, \ldots, t_K, t_j < t_{j+1} \in [0, T] \}$, our goal is to estimate the coefficients $\{\boldsymbol{\phi}(t): 0 \leq t \leq T\}$ given the sequence $\mathbf{x}$. Let $\mathbf{B}_s = \{ B_{s, \ell}; \, \ell = 1, \ldots, q \}$ be a vector of length $q$ of which elements are the basis functions evaluated at time $s$. By applying the basis representation in (\ref{eq:basis_representation}), we denote $\boldsymbol{\phi}_s = \boldsymbol{\Phi} \mathbf{B}_s$ as the smooth function $\boldsymbol{\phi}(\cdot)$ evaluated at time $s$. Therefore, this estimation reduces to the task of estimating the $p \times q$ coefficient matrix $\boldsymbol{\Phi}$. A major obstacle in obtaining the maximum likelihood estimators of the parameters in Model \eqref{eq:VCERGM}, similar to that of fitting an ERGM, is that calculation of the normalizing constant in the denominator is computationally intractable. Although numerical approaches such as the Markov chain Monte Carlo method can be used to estimate $\boldsymbol{\Phi}$ for small networks \citep{hunter:2006, wilson2017stochastic}, the computational cost is prohibitive for moderate to large networks, let alone a sequence of networks. To alleviate the computational complexity, we exploit a maximum pseudo-likelihood approach, originally adapted for fitting the ERGM \citep{strauss1990pseudolikelihood, van2009framework, wasserman:1996}. We show that the maximum pseudo-likelihood estimator (MPLE) for the VCERGM can be efficiently obtained via maximum likelihood estimation of a logistic regression model. Below we describe the estimation procedure in more detail.

% Time-varying network size
Without loss of generality, we assume that the numbers of nodes in different networks are the same over time, i.e., $n_s \equiv n$ for all $s$. In the case where $n_s$ varies over time, one can simply use the normalized network statistics described in Section \ref{sec:model}. For each observed time point $s = t_1, \ldots, t_K$, let $X^{s}_{ij}$ denote the binary random variable that describes whether or not there is an edge between node $i$ and node $j$ at time $s$. Furthermore, let $\mathbf{X}^s_{-(ij)}$ collection of ${n \choose 2} - 1$ binary random variables that describe whether or not there is an edge between all other pairs of nodes other than the node pair $i$ and $j$. For each observed time point $s = t_1, \ldots, t_K$, assume the conditional independence between edges. The marginal pseudo-likelihood function of $\boldsymbol{\Phi}$ given $x_s$ at time $s$ is defined as
\begin{equation}\label{eq:PL_marginal}
	\text{PL}(\boldsymbol{\Phi} | x_s) = \prod_{i,j \in [n]} \mathbb{P}(X_{ij}^s = x_{ij}^s | \mathbf{X}^{s}_{-(ij)} = \mathbf{x}^s_{-(ij)}).
\end{equation}

% {\color{red} GL: I still have some reservation about the notations. The small x represents a network (an $n\times n$ binary matrix), a sequence of networks, or the realization of a binary random vector in different contexts. Given the complexity of the notations, it may be helpful to have a paragraph introducing the notations in the introduction.} #Note: there is no longer any instance of using this as a binary random vector. Just keep it a sequence. So bold means sequence, and non-bold means matrix.

{\noindent Subsequently, the marginally independent composite pseudo likelihood of model \eqref{eq:VCERGM} is}
\begin{equation}\label{eq:PL_joint}
	\text{PL}(\boldsymbol{\Phi} | \mathbf{x}) = \prod_{s = t_1}^{t_K} \prod_{i,j \in [n]} \mathbb{P}(X_{ij}^s = x_{ij}^s | \mathbf{X}^{s}_{-(ij)} = \mathbf{x}^s_{-(ij)}). \nonumber
\end{equation}
The MPLE $\widehat{\boldsymbol{\Phi}}$ is obtained by maximizing $\text{PL}(\boldsymbol{\Phi} | \mathbf{x})$.
%\begin{equation}\label{eq:MPLE}
%	\widehat{\boldsymbol{\Phi}} = \text{argmax}_{\Phi}\{\text{PL}(\boldsymbol{\Phi} | \mathbf{x})\}.
%\end{equation}
%Our goal then is to calculate the MPLE $\widehat{\boldsymbol{\Phi}} = \text{argmax}_{\Phi}\{\text{PL}(\boldsymbol{\Phi} | \mathbf{x})\}$.

Let $x_{s, ij}^+$ denote an $n \times n$ matrix whose $(i, j)$th entry $x^s_{ij} = 1$ and let $x_{s, ij}^-$ be the same $n \times n$ matrix except $x^s_{ij} = 0$. Define $\boldsymbol{\Delta}^s_{ij} = \mathbf{h}(x_{s, ij}^+) - \mathbf{h}(x_{s, ij}^-)$ as the vector describing the element-wise difference in the network statistics when $x^s_{ij}$ changes from 0 to 1. One can readily show that for each $s= t_1, \ldots, t_K$, the following relationship holds for all $i,j \in [n]$: 

\begin{align} \label{eq:Logit}
\text{logit}\left\{ \mathbb{P}(X^{s}_{ij} = 1 | \mathbf{X}^{s}_{-(ij)} = \mathbf{x}^s_{-(ij)}) \right\} &= \log \left\{ \frac{\mathbb{P}(X^{s}_{ij} = 1 | \mathbf{X}^{s}_{-(ij)} = \mathbf{x}^s_{-(ij)})}{\mathbb{P}(X^{s}_{ij} = 0 | \mathbf{X}^{s}_{-(ij)} = \mathbf{x}^s_{-(ij)})} \right\} \nonumber \\
& = \log \left[ \exp\{\boldsymbol{\phi}(t)^T(\mathbf{h}(x_{s, ij}^+) - \mathbf{h}(x_{s, ij}^-))\} \right] \nonumber\\
& = \boldsymbol{\phi}_s^T \boldsymbol{\Delta}^s_{ij}
\end{align}

Let $Y_{ij}^s = \text{logit}\{ \mathbb{P}(X^{s}_{ij} = 1 | \mathbf{X}^{s}_{-(ij)} = \mathbf{x}^s_{-(ij)})\}$ and let $\mathbf{Y}_s = (Y_{11}^s, Y_{12}^s, \ldots, Y_{nn}^s)^T$. Similarly, define $\boldsymbol{\Delta}_s = (\boldsymbol{\Delta}^s_{11}, \boldsymbol{\Delta}^s_{12}, \ldots \boldsymbol{\Delta}^s_{nn})$ as the $p \times \binom{n}{2}$ matrix whose $r$th row contains the change in the $r$th network statistic when each edge changes from 0 to 1. Let $\text{vec}(X)$ be the operator that stacks the columns of $X$ into a column vector, and let $\otimes$ represent the Kronecker product operator. Combining (\ref{eq:basis_representation}) and (\ref{eq:Logit}) yields
\begin{equation}\label{eq:vector}
	\mathbf{Y}_s = \boldsymbol{\Delta}_s^T\boldsymbol{\Phi}\mathbf{B}_s = (\mathbf{B}_s \otimes \boldsymbol{\Delta}_s)^T~\text{vec}(\boldsymbol{\Phi}), \hskip 1pc \text{$s = t_1, \ldots, t_K$.}
\end{equation}
Let $\mathbf{Y} = (\mathbf{Y}_{t_1}, \cdots,\mathbf{Y}_{t_K})^T$, and define the $K \binom{n}{2} \times pq$ design matrix $\mathbf{H}$ as
\begin{equation}
\mathbf{H}
= \begin{pmatrix}
\mathbf{B}_{t_1} \otimes \boldsymbol{\Delta}_{t_1}  \\
\vdots \\
\mathbf{B}_{t_K} \otimes \boldsymbol{\Delta}_{t_K}  \\
\end{pmatrix}. \nonumber
\end{equation}
The relationship in (\ref{eq:vector}) connects to a logistic regression where $\mathbf{H}$ represents a design matrix with its coefficient $\text{vec}(\boldsymbol{\Phi})$.
% $\mathbf{Y} = \mathbf{H} ~ \text{vec}(\boldsymbol{\Phi})$.
%\begin{equation}\label{eq:logistic_regression}
%	\mathbf{Y} = \mathbf{H} ~ \text{vec}(\boldsymbol{\Phi}).
%\end{equation}
In \cite{strauss1990pseudolikelihood}, it was shown that maximizing the pseudo-likelihood PL($\boldsymbol{\Phi} | x_s)$ in (\ref{eq:PL_marginal}) is equivalent to finding the maximum likelihood estimator (MLE) of $\boldsymbol{\Phi}$ in the logistic regression model given in (\ref{eq:Logit}) with independent entries $X_{ij}^{s}$. It follows from the independence of $\mathbf{X}_{s}$ and $\mathbf{X}_{s'}$ for $s \neq s'$ that maximizing PL($\boldsymbol{\Phi} | \mathbf{x}$) is equivalent to calculating the MLE of $\boldsymbol{\Phi}$ in the logistic regression model $\mathbf{Y} = \mathbf{H}^T \text{vec}(\boldsymbol{\Phi})$ treating $\{X_{ij}^s: i,j \in [n], s = t_1, \ldots, t_K\}$ as mutually independent variables. % Thus the MPLE $\widehat{\boldsymbol{\Phi}}$ is equivalent to the MLE obtained from fitting $\mathbf{Y} = \mathbf{H}^T \text{vec}(\boldsymbol{\Phi})$ for independent entries.

\subsection{Penalized Logistic Regression}\label{subsec:penalized}
% Penalization
To obtain smooth estimates of the time-varying coefficients $\boldsymbol{\phi}(t)$, we further consider a roughness penalty on the coefficients of the basis functions \citep[see][for example]{eilers1996flexible,hoover1998nonparametric}. 
A commonly used penalty, which we use throughout this paper, is the integrated squared second derivative defined for $k$th row of $\boldsymbol{\Phi}$, denoted as $\boldsymbol{\Phi}_{(k)}$, as
$$\mathcal{P}(\boldsymbol{\Phi}_{(k)}) =  \int \{ D^2 \phi_k(u) \} ^2 \, du  = \boldsymbol{\Phi}_{(k)}^T \, \boldsymbol{\Omega} \, \boldsymbol{\Phi}_{(k)}$$
\noindent where a smoothness matrix $\boldsymbol{\Omega}$ in this case can be specified as
$$
\boldsymbol{\Omega} = \Bigl\{\Omega_{ij} = \int \{D^2 B_i(u) \} \{ D^2 B_j(u) \} \, du \,; \, \, i, \, j = 1, \ldots, q \Bigr\}.
$$
\noindent For networks observed at discrete time points $t_1, \ldots, t_K$, the $(i, j)$th element of $\boldsymbol{\Omega}$ is 
$$
\Omega_{ij} = \sum_{s=t_1}^{t_K} \{ D^2 B_i (s) \} \{ D^2 B_j (s) \}, \quad i, \, j = 1, \ldots, q.
$$

For more examples of possible penalties, see the Chapter 5 in \citet{ramsay2006functional}. As the same collection of basis functions are used to express $\phi_k(t)$, $k = 1, \ldots, p$, via basis representation, we impose the same $\boldsymbol{\Omega}$ on all $\phi_k(t)$. Consequently, we add the penalty term
\begin{equation}\label{eq:penalty}
\mathcal{P}_{\boldsymbol{\Omega}}(\boldsymbol{\Phi})=\sum_{k=1}^p \, \boldsymbol{\Phi}_{(k)}^T \, \boldsymbol{\Omega} \, \boldsymbol{\Phi}_{(k)} = \text{vec}(\boldsymbol{\Phi})^T (\boldsymbol{\Omega} \otimes \mathbf{I}_p)\text{vec}(\boldsymbol{\Phi}) \nonumber
\end{equation}
\noindent to the logistic log pseudo likelihood function. In conclusion, we calculate the penalized pseudo-likelihood estimator $\widehat{\boldsymbol{\Phi}}_{\Omega}$ by maximizing the following penalized log likelihood with tuning parameter $\lambda$:
\begin{equation}\label{eq:final}
\mathbf{y}^T\mathbf{H}\text{vec}(\boldsymbol{\Phi}) -\log[1+\exp\{\mathbf{H}\text{vec}(\boldsymbol{\Phi})\}] - \lambda \mathcal{P}_{\boldsymbol{\Omega}}(\boldsymbol{\Phi}).
\end{equation}

To fit (\ref{eq:final}), we implement an iteratively reweighted least squares (IRLS) algorithm. A detailed description of this procedure is available in supplementary materials. %Appendix \ref{appendix:IRLS}.

%%%%%%%%%%%%%%%%
% Section: Hypothesis testing %
%%%%%%%%%%%%%%%%

\section{Testing for Heterogeneity}\label{sec:hypothesis}

% Hypothesis testing
A key assumption of the VCERGM is that the effects of a specified collection of statistics vary through time. This assumption reflects heterogeneity in an observed sequence of graphs $\mathbf{x}$, and provides intuition as to whether or not summaries of $\mathbf{x}$ can be treated in aggregate. One can formally test for heterogeneity in $\mathbf{x}$ using a likelihood ratio test (LRT), which we will now describe.

We begin with the preconceived notion that $\mathbf{x}$ is homogeneous, namely that the coefficients $\boldsymbol{\phi}(t)$ under model (\ref{eq:VCERGM}) are fixed as constants over time. This serves as the null model, under which the VCERGM is equivalent to fitting independent and identically distributed ERGMs. With fixed constants $\phi_1^0, \ldots, \phi_p^0$, the null hypothesis corresponding to a homogeneous sequence of graphs can be written as
\begin{equation}\label{hyp_1}
\text{H}_{0}: \phi_1(t) = \phi_1^0, \, \, \ldots, \, \, \phi_p(t) = \phi_p^0.
\end{equation}

Setting the function $\phi_k (t) = \phi_k^0$ is equivalent to writing $\Phi_{k\ell} = \phi^0_k$ for all $\ell = 1, \ldots q$. As a result, the null hypothesis in (\ref{hyp_1}) can be expressed more succinctly as
\begin{equation}\label{hyp_2}
\text{H}_{0}: \boldsymbol{\Phi} = \boldsymbol{\Phi}^0 = (\phi_1^0, \ldots, \phi_p^0)^{^T} \times \mathbf{1}_q^{^T}, \nonumber
\end{equation}

\noindent where $\mathbf{1}_q$ is length $q$ vector of 1's. The coefficients under the null hypothesis are the restricted form of the VCERGM where the basis coefficients for each network statistic are constants for all $q$ basis functions. 

% In the model (\ref{eq:VCERGM}), setting the function $\phi_k (t)$ as a flat function of constant $\phi_k^0$ is equivalent to writing $\phi_k (t) = \phi_k^0 \sum_{l = 1}^q B_l (t), k = 1, \ldots p$. As a result, the null hypothesis can be written with respect to the coefficient matrix $\boldsymbol{\Phi}$ as
%$$
%\text{H}_{0}: \boldsymbol{\Phi} = \boldsymbol{\Phi}^0 = (\phi_1^0, \ldots, \phi_p^0) \times \mathbf{1}_q^{^T}
%$$

% Likelihood ratio test
The likelihood ratio test (LRT) is commonly used for conducting the test for heterogeneity in varying-coefficient models \citep{cai2000efficient, fan2001generalized, fan2000simultaneous, fan2008statistical}. Due to the dependence between entries in each graph, we utilize a pseudo likelihood ratio test (pLRT) \citep{staicu2014likelihood}. As previously emphasized in Section \ref{sec:estimation}, the joint pseudo likelihood consists of the distribution of $X_{ij}^s$ given the rest of the data $\mathbf{X}_{-(ij)}^s$ for all $i, j \in [n], s = t_1, \ldots t_K$. Furthermore, maximizing the pseudo likelihood simplifies the estimation process as fitting a logistic regression. Namely, with observed networks $\mathbf{x} = \{ x_{s}: s = t_1, \ldots, t_K \}$ with $n$ nodes, the pLRT compares the pseudo log likelihood function below under the null and alternative hypotheses:
%\begin{eqnarray}
%L(\boldsymbol{\Phi}) & = & \prod_{t=1}^T \prod_{i \in [n], j \in [n]} \Biggl( \frac{\exp[\mathbf{B}_{t}^{'} \boldsymbol{\Phi}^{'} \mathbf{g}(X_{ij}^{t} = 1, X_{-(ij)}^t) ]}{\exp[\mathbf{B}_{t}^{'} \boldsymbol{\Phi}^{'} \mathbf{g}(X_{ij}^{t} = 0, X_{-(ij)}^t) ]  + \exp[\mathbf{B}_{t}^{'} \boldsymbol{\Phi}^{'} \mathbf{g}(X_{ij}^{t} = 1, X_{-(ij)}^t) ] ] } \Biggr)^{X_{ij}^{t}} \\
%& & \times \Biggl( \frac{\exp[\mathbf{B}_{t}^{'} \boldsymbol{\Phi}^{'} \mathbf{g}(X_{i,}^{t} = 0, X_{-(ij)}^t) ]}{\exp[\mathbf{B}_{t}^{'} \boldsymbol{\Phi}^{'} \mathbf{g}(X_{ij}^{t} = 0, X_{-(ij)}^t) ]  + \exp[\mathbf{B}_{t}^{'} \boldsymbol{\Phi}^{'} \mathbf{g}(X_{ij}^{t} = 1, X_{-(ij)}^t) ] ] } \Biggr)^{1 - X_{ij}^{t}} \\
%\end{eqnarray}
%\begin{eqnarray}
%l(\boldsymbol{\Phi}|\mathbf{X}) = \sum_{s = t_1}^{t_K} \sum_{i \in [n], j \in [n]} \Bigl[  X_{ij}^{s} \log \Bigl( \frac{\exp[\mathbf{B}_{s}^{'} \boldsymbol{\Phi}^{'} \mathbf{h}_s(X_{ij}^{s} = 1, X_{-(ij)}^s) ]}{\exp[\mathbf{B}_{s}^{'} \boldsymbol{\Phi}^{'} \mathbf{h}_s(X_{ij}^{s} = 0, X_{-(ij)}^s) ] ]} \Bigr) - \log \Bigl(1 + \frac{\exp[\mathbf{B}_{s}^{'} \boldsymbol{\Phi}^{'} \mathbf{h}_s(X_{ij}^{s} = 1, X_{-(ij)}^s) ]}{\exp[\mathbf{B}_{s}^{'} \boldsymbol{\Phi}^{'} \mathbf{h}_s(X_{ij}^{s} = 0, X_{-(ij)}^s) ]} \Bigr) \Bigr]
%\end{eqnarray}
\begin{eqnarray*}
\log \text{PL} (\boldsymbol{\Phi}|\mathbf{x}) & = & \sum_{s = t_1}^{t_K} \sum_{i, j \in [n]} \log \{\mathbb{P}(X_{ij}^s = x_{ij}^s | \mathbf{X}_{-(ij)}^s = \mathbf{x}_{-(ij)}^s )\} \nonumber \\ 
& = & \sum_{s = t_1}^{t_K} \sum_{i, j \in [n]}  \Bigl[  x_{ij}^{s} \mathbf{B}_{s}^{T} \boldsymbol{\Phi}^{T} \boldsymbol{\Delta}_{ij}^s - \log \{ 1 +  \exp ( \mathbf{B}_{s}^{T} \boldsymbol{\Phi}^{T} \boldsymbol{\Delta}_{ij}^s ) \} \Bigr].
\end{eqnarray*}

Let $\hat{\boldsymbol{\Phi}}_{\text{H0}}$ and $\hat{\boldsymbol{\Phi}}_{\text{H1}}$ be the estimates of $\boldsymbol{\Phi}$ under the null and alternative hypotheses. The estimate  $\hat{\boldsymbol{\Phi}}_{\text{H1}}$ can be calculated by fitting the VCERGM specified in (\ref{eq:VCERGM}) and $\hat{\boldsymbol{\Phi}}_{\text{H0}}$ is the estimate from the VERCM with a restriction of constant basis coefficients. Accordingly, let $\log \text{PL}(\hat{\boldsymbol{\Phi}}_{\text{H0}}|\mathbf{x})$ and $\log \text{PL}(\hat{\boldsymbol{\Phi}}_{\text{H1}}|\mathbf{x})$ denote the pseudo log likelihood functions under the null and alternative, respectively. Then, the test statistic is
\begin{eqnarray}\label{eq:pLRT}
T & = & 2 \{ \log \text{PL}(\hat{\boldsymbol{\Phi}}_{\text{H1}}|\mathbf{x}) - \log \text{PL}(\hat{\boldsymbol{\Phi}}_{\text{H0}}|\mathbf{x})\} \nonumber \\
& = & 2 \sum_{s = t_1}^{t_K} \sum_{i, j \in [n]} \Bigl[ x_{ij}^{s} B_{s}^{T} (\hat{\boldsymbol{\Phi}}_{\text{H1}} - \hat{\boldsymbol{\Phi}}_{\text{H0}})^{T} \boldsymbol{\Delta}_{ij}^s+ \log \Bigl\{ \frac{1 + \exp ( B_{s}^{T}  \hat{\boldsymbol{\Phi}}_{\text{H0}}^{T} \boldsymbol{\Delta}_{ij}^s ) }{1 + \exp ( B_{s}^{T} \hat{\boldsymbol{\Phi}}_{\text{H1}}^{T} \boldsymbol{\Delta}_{ij}^s ) }  \Bigr\} \Bigr]. 
\end{eqnarray}

We reject the null hypothesis when $T > C_\alpha$ where $C_\alpha$ is the critical value of the test with significance level $\alpha$. In general, there exist two strategies commonly used to compute $C_\alpha$. The first approach is to apply Wilks phenomenon for deriving the asymptotic null distribution of the test statistic \citep{boucheron2011high, fan2001generalized}. Assumptions required to validate the asymptotic chi-squared distribution of the test statistic are presented in \cite{fan2001generalized}. Even though Wilks phenomenon is readily applicable to our case, calculating the degrees of freedom for asymptotic chi-squared distribution is difficult, and the chi-squared distribution may not be appropriate when the network size is relatively small. Another approach, which is preferable for moderate network size, involves generating bootstrap samples to construct the null distribution of $T$ \citep{cai2000efficient, fan2008statistical, huang2002varying}.
%The method of bootstrapping is found to be valid for logistic regression \citep{cai2000efficient}.
Analogous to the work in \cite{de2006generalized, mclachlan1987bootstrapping, tekle2016power}, the steps of obtaining the critical value $C_\alpha$ or calculating the $p$-value with parametric bootstrapping can be described as follows. 
For a large value of $B$, the test statistics (\ref{eq:pLRT}) calculated based on $B$ bootstrap samples successfully represent the null distribution of $T$.

\begin{enumerate}
\item Create $B$ bootstrap samples. For each bootstrap, indexed by $b = 1, \ldots, B$,
$\mathbf{x}^{*(b)} = \{ x_s^{*(b)}: s =t_1, \ldots, t_K \}$ is a sample from $\mathbb{P}(\mathbf{X} | \hat{\boldsymbol{\Phi}}_{\text{H0}})$.

% $\mathbf{x}^{*(b)} = \{ x_s^{*(b)}: s =t_1, \ldots, t_K \}$ is a random sample by sampling $\mathbf{x} = \{ x_s: s =t_1, \ldots, t_k \}$ with replacement. Namely, for all $s$, the $x_s^{*(b)}$ is uniformly distributed on the set $\mathbf{x}$ (i.e. $Pr(x_s^{*(b)} = x_s) = 1 / K$, $\forall s$).

\item For each bootstrap sample $\mathbf{x}^{*(b)}$, estimate $\boldsymbol{\Phi}$ under the null and alternative hypotheses and denote them as $\hat{\boldsymbol{\Phi}}_{\text{H0}}^{*(b)}$ and $\hat{\boldsymbol{\Phi}}_{\text{H1}}^{*(b)}$, respectively.

\item Calculate the test statistic for each bootstrap sample as
$$T^{*(b)} = 2 \{ \log \text{PL}(\hat{\boldsymbol{\Phi}}_{\text{H1}}^{*(b)}|\mathbf{x}^{*(b)}) - \log \text{PL}(\hat{\boldsymbol{\Phi}}_{\text{H0}}^{*(b)}|\mathbf{x}^{*(b)}) \}, \quad b = 1, \ldots, B.$$

\item The critical value $C_\alpha$ is determined as the $(1-\alpha)$th quantile of $(T^{*(1)}, \ldots, T^{*(B)})$. The $p$-value is the proportion of times that the bootstrap test statistic values exceed the observed test statistic $T$. Define an indicator function $I(A)$ which takes a value of 1 if $A$ is true and 0 otherwise. Then the $p$-value can be written as
$$
p{\text{-value}} = \frac{\sum_{b=1}^B I(T < T^{*(b)})}{B}.
$$
\end{enumerate}

The $p$-value is then used to determine whether or not to reject the null hypothesis. For values below a specified significance value, $\alpha$, one rejects the null hypothesis in (\ref{hyp_1}) and decides that the sequence of networks does exhibit heterogeneity in its parameters. In our applications below, we choose $\alpha = 0.05$ when evaluating any hypothesis test. 
%Under the null, the test statistic $T$ asymptotically follows $\chi^2$-distribution with degree of freedom $d_1 - d_0$ where $d_0$ and $d_1$ are the number of parameters under the null and alternative hypotheses, respectively.

%%%%%%%%%%%%%%%%
% Section: Simulation Study   %
%%%%%%%%%%%%%%%%

\section{Simulation Study}\label{sec:simulations}

The goal of our simulation study is two-fold: (i) to evaluate the power of the hypothesis testing procedure described in Section \ref{sec:hypothesis}, and (ii) to assess the goodness of fit of the VCERGM on dynamic networks with various magnitudes of temporal heterogeneity. In Section \ref{subsec:sim1}, we evaluate the sensitivity of the hypothesis test in (\ref{hyp_1}) for detecting temporal heterogeneity in a sequence of networks with fluctuating parameters using both the bootstrap procedure and the chi-squared approximation. Section \ref{subsec:sim2} assesses the performance of the VCERGM under various varying-coefficient specifications. 
%We investigate four different specifications for $\boldsymbol{\phi}(t)$ and 
We compare the performance of the VCERGM with other competing methods using measurements of bias, variance, and robustness. We further investigate how the VCERGM performs when the networks are observed at unequally spaced time points due to missing networks.

\subsection{Power Evaluation for Testing Heterogeneity}\label{subsec:sim1}

We first investigate the power of the hypothesis test for heterogeneity that we introduce in Section \ref{sec:hypothesis}. To do so, we investigate both Type I and Type II errors of the test on dynamic networks over various magnitudes of temporal heterogeneity. We simulate 100 sequences of dynamic networks $\mathbf{x} = \{ \mathbf{x}_1, \ldots, \mathbf{x}_{100} \}$, where each sequence $\mathbf{x}_{w} = \{ x_{w, 1}, \ldots, x_{w, K} \}$, contains $K$ networks with 30 nodes observed at equally-spaced times $t_1, \ldots, t_K$ under the VCERGM that models the temporal contributions of the edge density statistic. We set the coefficient on the edge density term, $\phi(t)$, to be a sinusoidal curve with amplitude $M$ and period $T$. In particular, we model
$$
\phi(t) = M \sin \Bigl(\frac{2 \pi t}{T} \Bigr), \quad t \in [0, T].
$$
We vary the number of observed time points $K$ from 10 to 100, and the amplitude $M$ from 0 to 0.3 in increments of 0.05. For each value of $K$ and $M$, we calculate the proportion of cases that we reject the null hypothesis at a $\alpha = 0.05$ level out of the 100 simulated dynamic network sequences. Table \ref{table:HypothesisSimulation} reports these proportions when using the bootstrap procedure as well as the chi-squared approximation. 

% and calculate the  Greater values of $K$ indicate the presence of more time points observed in the domain of $[0, T]$. When $M$ is bigger, the true $\phi(t)$ fluctuates within greater range.
%
% We set a cycle of sinusoidal curve as $\phi(t)$. Namely, the true $\phi(t)$ with signal size $M$ is defined as
% That is, for the same signal size, greater $K$ implies finer grid $(0 = t_1 < \ldots < t_K = T)$ to observe the true $\phi(t)$. Table \ref{table:HypothesisSimulation} shows the proportion of cases where we reject the null out of 100 simulated sets of networks.

When $M = 0$, $\phi(t)$ is a constant function and as a result the proportion of rejections in this case provides an estimate for the Type I error of each test. From Table \ref{table:HypothesisSimulation}, we see that both strategies obtain a Type I error at or below 0.05, as desired. For $M > 0$, the proportion of rejections provides an estimate of the power of the test. We see that for higher signal (larger $M$) and for a larger number of observed networks (larger $K$), we obtain a higher power, as expected. Across $K$, we see in general that the bootstrap procedure is consistently more powerful than the use of the chi-squared reference distribution for each amplitude value $M$. For $M > 0.25$ the power of both tests reaches 1, indicating that heterogeneity is successfully identified by both tests. These results suggest that both tests are powerful for large enough signal size, and that the bootstrap procedure outperforms the chi-squared approximation for small signal sizes (between $M = 0.05$ and $0.20$).

\begin{table}[htbp]
\caption{\label{table:HypothesisSimulation}{\bf Simulation results}: Proportion of cases that we reject the null hypothesis out of 100 simulations at the significance level of $\alpha = 0.05$. Bootstrap samples of size $B = 1000$ and chi-squared asymptotic distribution is used to make a decision for hypothesis testing.}
\centering
\resizebox{.75\columnwidth}{!}{%
%\fbox{% 
\begin{tabular}{l|rrrrr|rrrrr}
\toprule
\multirow{2}{*}{$M$} & \multicolumn{5}{c|}{Bootstrap} & \multicolumn{5}{c}{Chi-squared asymptotic} \\ 
\cline{2-11}
 & $K = 10$ & 30 & 50 & 70 & 100  & $K = 10$ & 30 & 50 & 70 & 100 \\
\hline
\hline
0 & 0.03 & 0.01 & 0.01 & 0.03 & 0 & 0 & 0 & 0.01 & 0.03 & 0 \\
\hline
0.05 & 0.15 & 0.36 & 0.46 & 0.59 & 0.71 & 0.01 & 0.23 & 0.36 & 0.52 & 0.66 \\
0.1 & 0.42 & 0.77 & 0.91 & 0.93 & 0.97 &  0.1 & 0.71 & 0.9 & 0.9 & 0.98 \\
0.15 & 0.74 & 0.98 & 1 & 1 & 0.99 & 0.35 & 0.97 & 1 & 1 & 1 \\
0.2 & 0.98 & 1 & 1 & 1 & 1 & 0.71 & 1 & 1 & 1 & 1 \\
0.25 & 1 & 1 & 1 & 1 & 1 & 0.91 & 1 & 1 & 1 & 1 \\
0.3 & 1 & 1 & 1 & 1 & 1 & 0.99 & 1 & 1 & 1 & 1 \\
% 0.5 & 1 & 1 & 1 & 1 & 1 & 1 & 1 & 1 & 1 & 1 \\
\bottomrule
\end{tabular}}
%}
\end{table}

\subsection{Estimation Performance}\label{subsec:sim2}

% Introduce the simulation setting
We now evaluate the performance of VCERGM to accurately estimate fluctuating parameters $\boldsymbol{\phi}(t)$, $t \in [0, T]$. We consider four different settings for $\boldsymbol{\phi}(t)$: (i) sinusoidal curve $\phi(t) = a \sin \{ (t + b)/c \} + d$ of varying amplitude $a$; (ii) quadratic curve $\phi(t) = a (t - T/2)^2 + b$ of varying strength $a$; (iii) dynamic \erdos random graph with probability $p$ of edges; and (iv) non-smooth (spiky) functions as a form of a sequence of random numbers with varying mean and standard deviation for normal distribution. For each setting of varying coefficients, we model the occurrence of graphs using the VCERGM with edge density and reciprocity statistics (see Table \ref{table:NetworkStatistics}). We simulate 100 dynamic sequences of directed graphs $\{ \mathbf{x}_1, \ldots, \mathbf{x}_{100} \}$ where each sequence $\mathbf{x}_w = \{ x_{w, 1}, \ldots, x_{w, 50} \}$ is observed at $K = 50$ equally-spaced time points. We assume that the network size remains constant through time and consider estimation with networks of three different sizes $n = 30, 50, 100$. Furthermore, we repeat (i)-(iv) with 1, 5, and 10 randomly chosen networks removed from the time series to evaluate the performance on dynamic networks with observations missing at random. 

% Competing method / Measurement for comparison
For each simulated dynamic network, we compare the VCERGM with two other dynamic network models. First, we fit cross-sectional ERGMs, where the ERGM in model (\ref{eq:ERGM}) is fit separately at each of the $K$ observed time points. As an alternative competitive method, we also develop an ad hoc 2-step procedure, which adapts an ad hoc smoothing procedure after fitting cross-sectional ERGMs for observed networks. To assess the performance of each method, we calculate the integrated absolute error (IAE) of the estimated coefficient curves. It measures the sum of point-wise absolute difference between estimated curve $\hat{\phi}(t)$ and true curve $\phi(t)$ at observed time points $t_1, \ldots, t_K$, namely
$$
IAE(\phi(t), \, \hat{\phi}(t))= \sum_{s=t_1}^{t_K} | \phi(s) - \hat{\phi}(s) | .
$$

The mean and standard deviation (SD) are calculated to evaluate the performance of our proposed method compared to cross-sectional ERGMs and ad hoc 2-step procedure. 
% They serve as measurements of bias and robustness of estimates from each method. 
We provide the summary of IAE for each method on dynamic networks with 30 nodes in Table \ref{table:Simulation30} with $(0, 1, 5, 10)$ missing networks.
%, and plot the estimated coefficient curves in Figure \ref{figure:Simulation2}.
Settings for the results are (i) sinusoidal curves with $(a, b, c, d) = (1, 20, 5, 1)$ (edges) and $(a, b, c, d) = (0.6, 20, 3, 0.4)$ (reciprocity); (ii) quadratic curves with $(a, b) = (1/25^2, 0)$ (edges) and $(a, b) = (-1/30^2, 0.5)$ (reciprocity); (iii) \erdos with $p_{edges} = 0.85$; (iv) a sequence of random numbers from $N(0, 1)$ (edges) and $N(1.5, 0.6)$ (reciprocity). 
%The results are robust across network size.
The performances of cross-sectional ERGMs, ad hoc 2-step procedure, and VCERGM become more comparable with larger network size. For results of $n = 50$ and $n = 100$ case, see Tables \ref{table:Simulation50} and \ref{table:Simulation100} in supplementary materials. %Appendix \ref{appendix:result}.

\begin{figure}[htbp]
\centering
\includegraphics[width = 0.9\textwidth, height = 0.35\textheight]{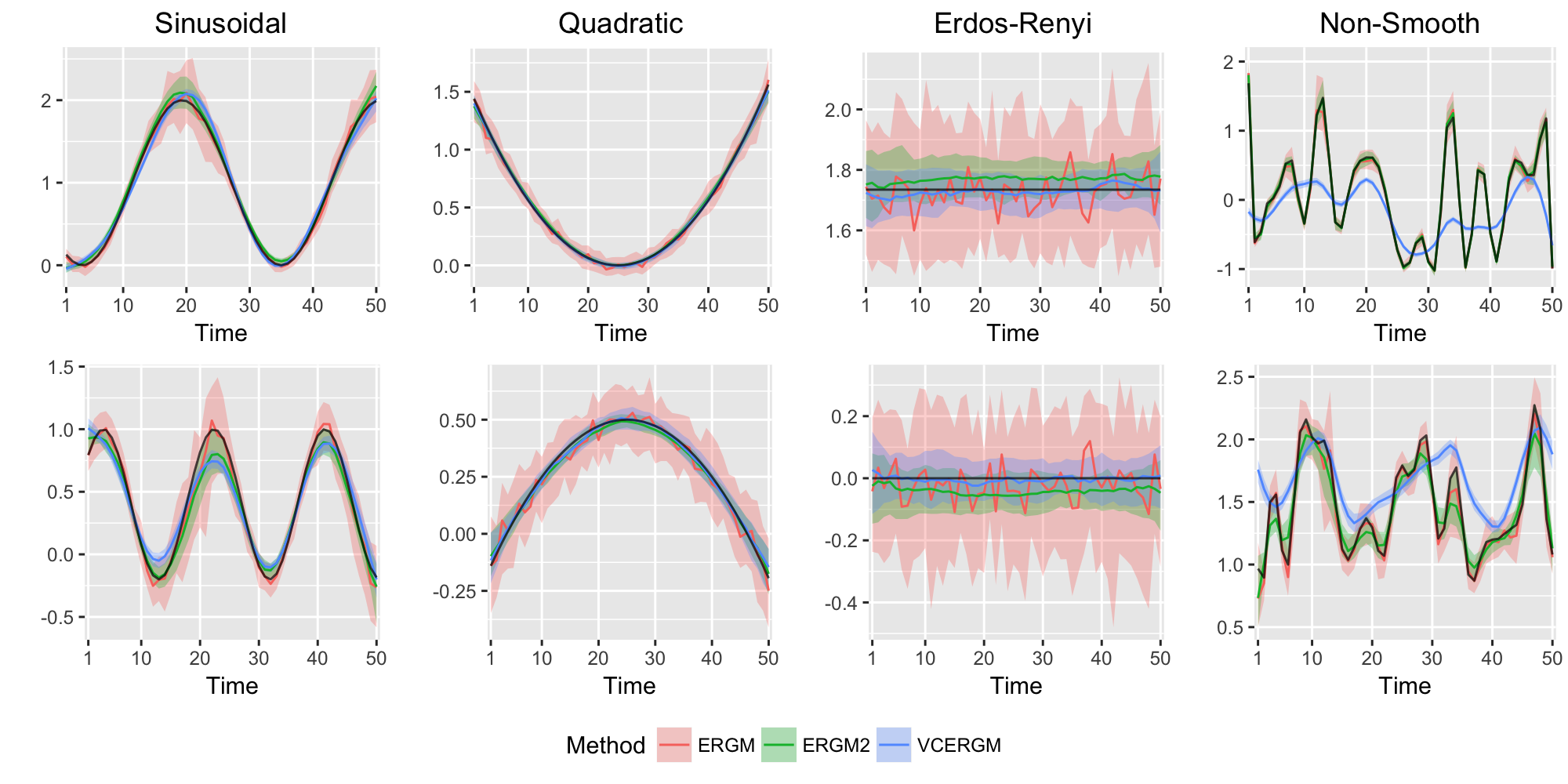}
\caption{{\bf Parameter estimates with 30 nodes}: Estimated parameters for edges (top) and reciprocity (bottom). Black line is the true $\boldsymbol{\phi}(t)$. Red (ERGM) is for cross-sectional ERGMs, green (ERGM2) is for ad hoc 2-step procedure, and blue (VCERGM) is for VCERGM. For each method, solid line indicates the average of 100 estimated curves and the shaded band illustrates the first and third quantiles.}
\label{figure:Simulation2}
\end{figure}

%{\color{red} Put the proportion of times that the dynamic sequence was deemed heterogeneous in the table! And talk about this.}
% Result / Figure
Without missing network, Figure \ref{figure:Simulation2} shows that cross-sectional ERGMs are more likely to introduce unexpected spikes or increased variability in estimating true $\boldsymbol{\phi}(t)$, compared to VCERGM. Overall, the VCERGM presents smaller deviation from true $\boldsymbol{\phi}(t)$ with smaller variability compared to cross-sectional ERGMs and ad hoc 2-step procedure. 
%Consistent with Table \ref{table:Simulation}, the VCERGM shows significantly less variability in parameter estimation in Figure \ref{figure:Simulation2} compared to cross-sectional ERGMs and ad hoc 2-step procedure. 
In the first three settings, the VCERGM performs better than cross-sectional ERGMs and ad hoc 2-step procedure with smaller IAE. In case of non-smooth functions, with true $\boldsymbol{\phi}(t)$ not smooth but wiggly, the ad hoc 2-step procedure shows better performance than the VCERGM with respect to IAE. Both Table \ref{table:Simulation30} and Figure \ref{figure:Simulation2} indicate that the VCERGM potentially misses random wiggles, which causes greater bias on average compared to cross-sectional ERGMs. Regardless, overall trend that the true non-smooth $\boldsymbol{\phi}(t)$ presents is fairly well estimated by the VCERGM and the variability of the estimates is still smaller for the VERGM compared to cross-sectional ERGMs. Overall, the performance of ad hoc 2-step procedure and VCERGM is comparable, but the VCERGM is more principled in terms of incorporating time-varying coefficients in the modeling step.
For all four settings, the VCERGM is computationally more efficient than the cross-sectional ERGMs.
% (ERGM 463 seconds; VCERGM 251 seconds on average for 100 sequences of networks). 
We conduct an additional simulation study specifically tailored to compare the computing time between methods and the results are presented in Table \ref{table:Computing}.

% Results with missing networks
% We provide the IAE of each method in Table \ref{table:Simulation30} and show the coefficient estimates in Figure \ref{figure:Simulation3}. 
When there exist missing networks, cross-sectional ERGMs are no longer available to provide the estimates at unobserved time points. %Therefore, the IAE is calculated based on observed time points only. 
Therefore, the IAE is calculated only for ad hoc 2-step procedure and VCERGM. Notably, the performance of the VCERGM remains stable across each number of missing networks. Cross-sectional ERGMs and the 2-step approach, on the other hand, suffer more than the VCERGM in the case of missing networks. Indeed, as shown in Table \ref{table:Simulation30}, the VCERGM outperforms these competitive methods in the case that observations are missing and is better able to capture the true coefficient curve in these cases. 

\begin{table}[htbp]
\caption{\label{table:Simulation30} {\bf Simulation results with 30 nodes and (0, 1, 5, 10) missing networks}: Mean and standard deviation of the integrated absolute errors (IAE) for each method.}     
\centering
\resizebox{\columnwidth}{!}{
%\fbox{%
\begin{tabular}{lr|rrr|rrr}
\toprule
& \multirow{2}{*}{Missing} & \multicolumn{3}{c|}{Edges} & \multicolumn{3}{c}{Reciprocity} \\
\cline{3-8}
& & \multicolumn{1}{c}{ERGM} & \multicolumn{1}{c}{ERGM2} & \multicolumn{1}{c|}{VCERGM} & \multicolumn{1}{c}{ERGM} & \multicolumn{1}{c}{ERGM2} & \multicolumn{1}{c}{VCERGM} \\
\hline
{\bf Sinusoidal} & 0 & 12.92 (7.52) & 7.55 (8.26) & {\bf 5.58} (7.48) & 14.35 (4.21) & 8.01 (4.9) & {\bf 6.3} (1.91) \\ 
% & 1 &  12.81 (7.45) & 7.84 (8.26) & {\bf 6.42} (7.32) & 14.19 (4.21) & 8.34 (4.86) & {\bf 6.25} (1.91) \\
% & 5 & 11.91 (7.02) & 8.44 (8.29) & {\bf 6.13} (7.46) & 13.2 (4.23) & 9.08 (4.79) & {\bf 7.21} (1.75)  \\
% & 10 & {\bf 10.04} (6.14) & 12.15 (7.68) & 11.89 (6.75) & 11.24 (3.65) & 11.37 (3.92) & {\bf 9.47} (1.57) \\
 & 1 & & 7.84 (8.26) & {\bf 6.42} (7.32) &  & 8.34 (4.86) & {\bf 6.25} (1.91) \\
 & 5 &  & 8.44 (8.29) & {\bf 6.13} (7.46) &  & 9.08 (4.79) & {\bf 7.21} (1.75)  \\
 & 10 &  & 12.15 (7.68) & {\bf11.89} (6.75) &  & 11.37 (3.92) & {\bf 9.47} (1.57) \\
\hline
{\bf Quadratic} & 0 & 6.16 (3.93) & 2.75 (4.43) & {\bf 2.72} (4.41) & 8.16 (0.88) & {\bf 2.74} (1.47) & 3.11 (1.18) \\ 
% & 1 & 6.07 (3.89) & 2.77 (4.42) & {\bf 2.75} (4.4) & 8.02 (0.85) & {\bf 2.75} (1.49) & 3.12 (1.18) \\
% & 5 & 5.58 (3.64) & {\bf 2.89} (4.44) & 2.92 (4.41) & 7.3 (0.78) & {\bf 2.83} (1.44) & 3.25 (1.13) \\
% & 10 & 4.95 (3.18) & 3.7 (4.32) & {\bf 3.64} (4.29) & 6.51 (0.74) & {\bf 3.44} (1.55) & 3.68 (1.19) \\
 & 1 &  & 2.77 (4.42) & {\bf 2.75} (4.4) &  & {\bf 2.75} (1.49) & 3.12 (1.18) \\
 & 5 &  & {\bf 2.89} (4.44) & 2.92 (4.41) &  & {\bf 2.83} (1.44) & 3.25 (1.13) \\
 & 10 &  & 3.7 (4.32) & {\bf 3.64} (4.29) &  & {\bf 3.44} (1.55) & 3.68 (1.19) \\
\hline
{\bf \erdos} & 0 & 15.46 (9.83) & 6.74 (11.32) & {\bf 6.13} (11.34) & 15.45 (2.24) & 5.14 (2.43) & {\bf 4.58} (1.43) \\
% & 1 & 15.17 (9.65) & 6.74 (11.34) & {\bf 6.18} (11.35) & 15.15 (2.23) & 5.12 (2.51) & {\bf 4.62} (1.47) \\
% & 5 & 13.93 (8.84) & 6.8 (11.29) & {\bf 6.28} (11.28) & 13.93 (2.22) & 5.27 (2.5) & {\bf 4.76} (1.53) \\
% & 10 & 12.41 (7.91) & 6.98 (11.33) & {\bf 6.46} (11.27) & 12.43 (2.1) & 5.73 (2.98) & {\bf 5} (1.64) \\
 & 1 &  & 6.74 (11.34) & {\bf 6.18} (11.35) &  & 5.12 (2.51) & {\bf 4.62} (1.47) \\
 & 5 &  & 6.8 (11.29) & {\bf 6.28} (11.28) &  & 5.27 (2.5) & {\bf 4.76} (1.53) \\
 & 10 &  & 6.98 (11.33) & {\bf 6.46} (11.27) &  & 5.73 (2.98) & {\bf 5} (1.64) \\
\hline
{\bf Non-smooth} & 0 & 26.42 (3.37) & {\bf 25.53} (3.47) & 33.35 (1.92) & 20.95 (4.44) & {\bf 19.65} (4.17) & 23.28 (2.95) \\
% & 1 & {\bf 26.02} (3.36) & 26.21 (3.44) & 33.56 (1.89) & 20.77 (4.41) & {\bf 19.86} (4.17) & 23.37 (3.01) \\
% & 5 & {\bf 22.49} (3.32) & 26.71 (3.43) & 32.82 (1.93) & {\bf 18.45} (4.3) & 20.16 (4.2) & 23.56 (2.94) \\
% & 10 & {\bf 19.94} (2.96) & 34.79 (2.92) & 36.01 (1.76) & {\bf 16.38} (3.64) & 22.17 (3.9) & 23.38 (2.88) \\
 & 1 & &  {\bf 26.21} (3.44) & 33.56 (1.89) &  & {\bf 19.86} (4.17) & 23.37 (3.01) \\
 & 5 &  &  {\bf 26.71} (3.43) & 32.82 (1.93) &  &  {\bf 20.16} (4.2) & 23.56 (2.94) \\
 & 10 &  &  {\bf 34.79} (2.92) & 36.01 (1.76) & &  {\bf 22.17} (3.9) & 23.38 (2.88) \\
\bottomrule
\end{tabular}}
%}
\end{table}

% Computing time
In order to compare the computational efficiency, we vary the number of time points $K$ and record the computing time for VCERGM and cross-sectional ERGMs. Table \ref{table:Computing} summarizes the computing times of 500 simulated dynamic network sequences and displays how computing time changes as the number of time points $K$ changes. According to Table \ref{table:Computing}, the VCERGM takes significantly less time than cross-sectional ERGMs to complete the parameter estimation. Even if both VCERGM and cross-sectional ERGMs show linear increase in computing time, the rate of change is much smaller for VCERGM. Both methods entail $K$ separate steps to construct design matrix and response vector at each time point, but the cross-sectional ERGMs require $K$ separate MPLE steps while VCERGM only needs one estimation. In other words, the longer the time series of networks are, the more efficient VCERGM is compared to cross-sectional ERGMs.

\begin{table}[htbp]
\caption{\label{table:Computing} {\bf Computing Time}: Summary (Mean(SD)) of computing time (second) for dynamic networks with different number of time points $K$}
\centering
\resizebox{.75\columnwidth}{!}{%
%\fbox{%
\begin{tabular}{lccccc}
\toprule
& \multicolumn{5}{c}{Number of time points $K$} \\
\cline{2-6}
 & 20 & 40 & 60 & 80 & 100 \\
\hline
ERGM & 1.00 (0.05) & 2.09 (0.08) & 3.25 (0.13) & 4.25 (0.11) & 5.30 (0.12) \\ 
VCERGM  & 0.43 (0.03) & 0.82 (0.04) & 1.25 (0.07) & 1.63 (0.07) & 2.01 (0.06) \\ 
\bottomrule
\end{tabular}}
%}
\end{table}

% The TERGM shows the linear increase in computing time like cross-sectional ERGMs, but at a slower rate. It may be more efficient than cross-sectional ERGMs when the length of networks gets greater, but not as efficient as VCERGM.

%%%%%%%%%%%%%
% Section: Applications %
%%%%%%%%%%%%%

\section{Applications}\label{sec:applications}

In this section we apply the VCERGM to two case studies which portray differing amounts of temporal heterogeneity. First, we analyze how the co-voting patterns among U.S. Senators have changed through time. In this example, we analyze the effects of political affiliation (Republican or Democrat) on the likelihood of the voting networks. In our second example, we investigate the structural changes of the resting-state functional magnetic resonance imaging (rfMRI) records of healthy individuals. For each case study, we first test for temporal heterogeneity of any statistic included in the model. We fit the VCERGM on the two data sets and also fit cross-sectional ERGMs and ad hoc 2-step procedure for comparison. In the first example, we witness a clear evidence of temporal heterogeneity that the importance of political affiliation on the likelihood of the voting networks significantly changes over time. On the contrary, relatively stable rfMRI networks show little fluctuations through time and our method is shown to accommodate it successfully as well.

% and the VCERGM successfully captures the evolving structure of voting networks. 

\subsection{Political Voting Network} \label{subsec:political}

% Data introduction
We begin by analyzing the dynamic network that describes the co-voting patterns among U.S. Senators from 1867 (Congress 40) to 2015 (Congress 113). Three of the voting networks are shown in Figure \ref{figure:TemporalNetworks1}. This network was first investigated in \citet{moody2013portrait} and has been subsequently analyzed in \citet{wilson2016modeling}. The network is based off of the roll call voting data from  \url{http://voteview.com}, which contains the voting decision of each Senator (yay, nay, or abstain) for every bill brought to Congress.  We model the co-voting tendencies of the Senators using a dynamic network where nodes represent Senators and an edge is formed between two nodes if the two Senators vote concurrently (both yay or both nay) on at least 75\% of the bills to which they were both present. 

As shown in Figure \ref{figure:TemporalNetworks1}, there exist great fluctuations in network structure through time. Previous analyses in \citet{moody2013portrait, wilson2016modeling} have identified significant changes in the community structure of the network over time, and that this community structure is closely associated with political affiliation of the Senators. To account for these fluctuations, we incorporate a node-match statistic for political affiliation which counts the number of edges shared between Senators who have the same political affiliation. Furthermore, we include a statistic which models popularity (two-star) and clustering (triangle) of the Senator networks. See Table 1 for more details of these two statistics. We note that it is of separate interest to perform model selection for the VCERGM; however, we pursue this in future work. Here, we compare and contrast the results of a simple model when using several viable dynamic network models.

% Political: Estimate
\begin{figure}[htbp]
\centering
\includegraphics[width = \textwidth]{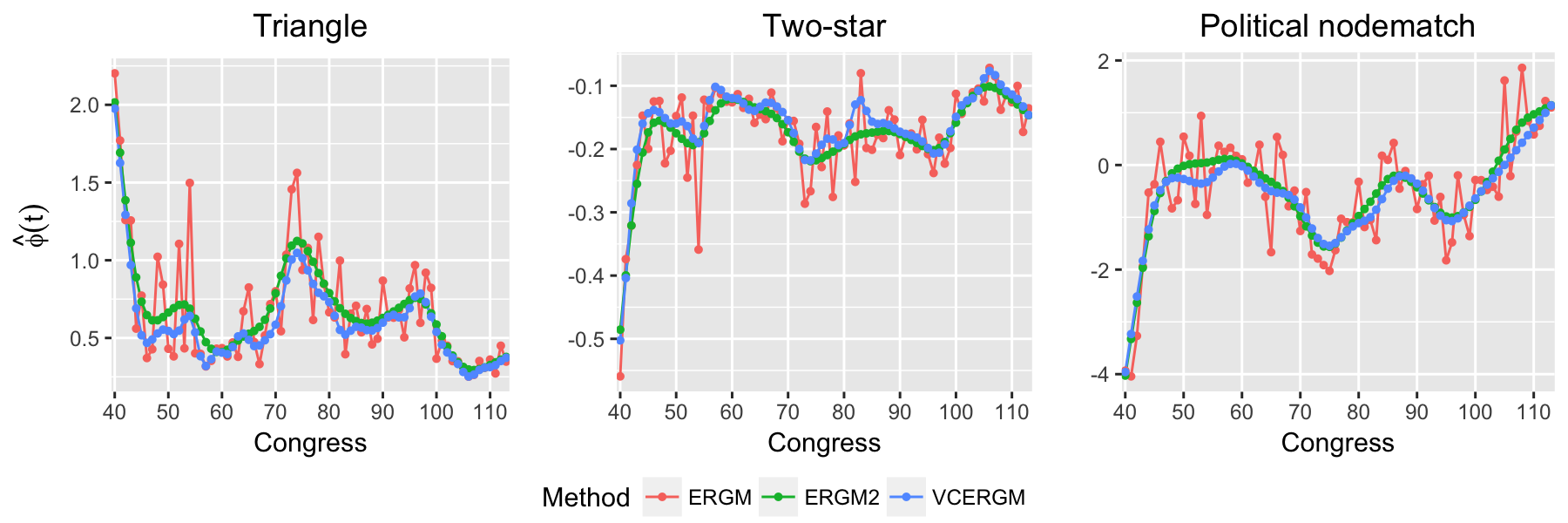}
\caption{{\bf Parameter estimates of political networks}:  Temporal heterogeneity is clearly presented for all three network statistics. Cross-sectional ERGMs (ERGM), ad hoc 2-step procedure (ERGM2) and VCERGM show similar estimates.}
\label{figure:PoliticalEstimate}
\end{figure}

% Result
The estimated parameters from i) cross-sectional ERGMs (ERGM), ii) ad hoc 2-step procedure (ERGM2) and iii) VCERGM are presented in Figure \ref{figure:PoliticalEstimate}. Notably, all three network statistics exhibit temporal heterogeneity. The asymptotic chi-squared $p$-value for testing heterogeneity is $<.001$.
%The test statistic for temporal heterogeneity is 13768.84 with asymptotic chi-squared $p$-value $<.001$. 
Consistent with our simulation results, the cross-sectional ERGMs exhibits spiky estimates, but the ad hoc smoothing recovers the lack of smoothness efficiently and produces similar estimates as the VCERGM. The political nodematch parameter estimate reveals an important trend in the political network. We see that the coefficient value for this term generally increases over time, indicating the increasing importance of political affiliation on the co-voting habits of the Senators throughout U.S. history. In particular since Congress 95, this coefficient has significantly increased. This finding matches the current theory of ``political polarization'' described in \cite{moody2013portrait}. Figure \ref{figure:PoliticalEstimate} also suggests that the triangle coefficient has decreased in magnitude suggesting that clustering has become less and less important over time.

%\begin{figure}[h!]
%\centering
%\includegraphics[width = 0.75\textwidth, height=0.25\textheight]{Figure/Political/GOF_Diff_lag=0_1.png}
%\caption{Political Networks: Goodness-of-Fit}
%\label{figure:PoliticalGOF}
%\end{figure}

\subsection{fMRI Dataset} \label{subsec:fMRI}

% Data source / preprocessing
We next analyze the structure of brain connectivity in the data provided by the WU-Minn Consortium Human Connectome Project (HCP). The dataset is available at \url{https://db.humanconnectome.org}. See \citet{van2012human} for an overview of data acquisition and analysis. The dataset includes the resting-state functional magnetic resonance imaging (rfMRI) of 500 subjects. For each subject, a 15-minute run of rfMRI is recorded. We set 47 local windows and calculate a partial precision matrix between 50 brain regions based on observations within each window. For a transition from precision matrices to sequence of dynamic networks, we define the edge density of a network as the proportion of edges in the network. Once the edge density is specified, the threshold of partial precision value can be determined to form an edge between the brain regions. With the edge density of 10\%, for example, the greatest 10\% of partial precision values would form edges. 

% Model setting
References \cite{simpson2011exponential} and \cite{simpson2012exponential} were among the first to fit ERGMs to functional brain networks. Their final model included network statistics such as geometrically weighted edge-wise shared partner (GWESP) and geometrically weighted non-edge-wise shared partner (GWNSP). For sake of comparison, we model our rfMRI networks with three network statistics: edges, triangle and two-star and compare i) cross-sectional ERGMs (ERGM), ii) ad hoc 2-step procedure (ERGM2) and iii) VCERGM. We note that model selection for the VCERGM (and other dynamic network models in general) is an important area for future research.

% Window = 200 / Density = 0.1
\begin{figure}[htbp]
\centering
\includegraphics[width = \textwidth]{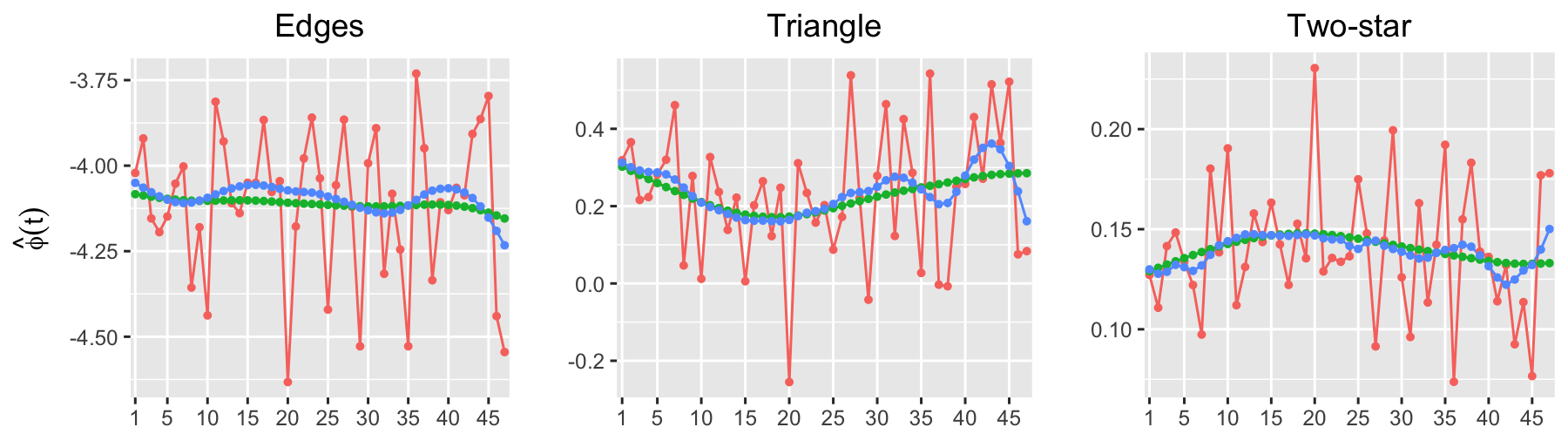}
\includegraphics[width = \textwidth]{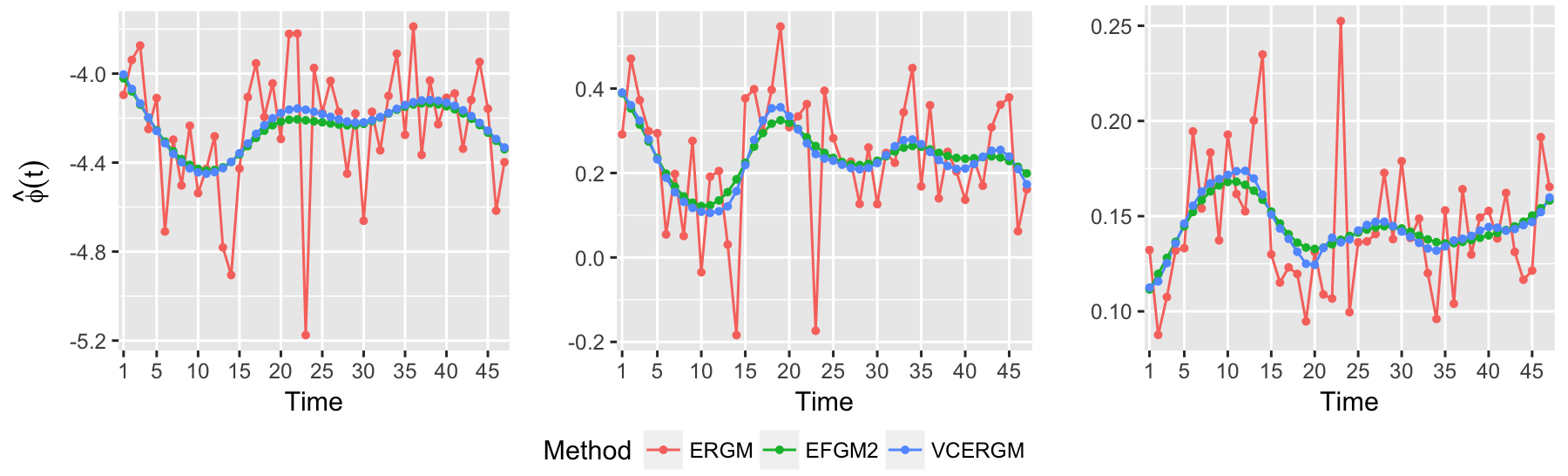}
\caption{{\bf Parameter estimates of fMRI networks}: Results of two randomly chosen individuals. For all three network statistics, one individual (second row; asymptotic chi-squared $p\text{-value}=0.999$) displays slightly more fluctuations than the other individual (first row; asymptotic chi-squared $p\text{-value}=1$). The ad hoc 2-step procedure and VCERGM show similar estimates.}
\label{figure:fMRIEstimate}
\end{figure}

% 33.38955 / 48.95913

% Result: Parameter estimate
Figure \ref{figure:fMRIEstimate} shows the results of two individuals from this study. As the data are the resting-state fMRI records, little fluctuation is expected in parameters over time. For both individuals, both ad hoc 2-step procedure and VCERGM provide estimates with a small range of fluctuation for all three network statistics. Overall, the ad hoc 2-step procedure and VCERGM provide relatively similar estimates, while both estimates cross the cross-sectional ERGM estimates. The estimates from cross-sectional ERGMs are extremely jagged that they may introduce inaccurate inference with regard to explaining the topological change in brain networks over time. 
%The ad hoc 2-step procedure seems to over flatten the estimates in a couple of cases that it may get rid of any meaningful small temporal variations in the dynamic networks. 
The VCERGM not only produces fairly static estimates but also captures small variations through time more sensitively than ad hoc 2-step procedure. Therefore, even with relatively stable dynamic networks, the VCERGM performs consistently well.

% Parameter 4. Result: Goodness-of-fit
%In terms of the goodness-of-fit, VCERGM is compared with cross-sectional ERGMs and ad hoc 2-step procedure. For each method, we simulate 100 networks by using the coefficient estimates at each time point. Then, we calculate the distance of distribution of 4 goodness-of-fit statistics between observed and simulated networks. Boxplots in Figure \ref{figure:fMRIGOF} illustrate the distribution of difference between VCERGM and the other two methods. The lower in the $y$-axis the boxplot is located, the more consistent VCERGM performs compared to ERGM and ad hoc 2-step procedure. {\color{red} Instead of boxplots, write a paragraph to describe what we see in terms of gof.}
%\begin{figure}[ht]
%\centering
%\includegraphics[width = 0.7\textwidth, height=0.2\textheight]{Figure/fMRI/GOF_Diff_lag=0_100307.png}
%\caption{fMRI Networks: Goodness-of-Fit}
%\label{figure:fMRIGOF}
%\end{figure}
% The more sparse a network gets, the better VCERGM performs compared to separate ERGMs.

%%%%%%%%%%%%%
% Section: Discussion   %
%%%%%%%%%%%%%

%{\color{red} If we move ERGM-TERGM relation to the supp, this section can be significantly shortened as well. You can remove the last three paragraphs, or move them to the supp}

\section{Discussion}\label{sec:discussion}
In this paper, we introduce varying-coefficient models for dynamic networks. In particular, we described the formulation and estimation of the VCERGM, a model that incorporates temporal changes in the coefficients of an exponential random graph family of models. We demonstrated the advantages of applying the VCERGM over competing methods through simulations and two dynamic network case studies. First, the VCERGM provides an intuitive explanation of how a network changes through time. Both the cross-sectional ERGMs and ad hoc 2-step procedure seemed to capture the temporal heterogeneity in a sense. However, by incorporating the temporal heterogeneity in the modeling step, the VCERGM provides a compact and meaningful model to formally explain the temporal structure of dynamic networks. Second, the VCERGM is robust to perturbations in observed temporal data. By imposing smoothness on the coefficients, we are able to provide robust estimates that are resistant to outliers and noise. Third, the VCERGM enables interpolation for missing networks through time. In practice, one can only observe a finite number of networks in a dynamic sequence, which may be observed in unequally spaced time increments. Estimates of the coefficients to the VCERGM can be evaluated at any time point in the domain and immediately interpreted as the impact of network statistics at that time point. By presenting the results with unequally-spaced networks, we illustrated how the varying-coefficients through time can be useful especially in terms of interpolation. 

Our work provides several avenues for future research. First, it is important to consider the evaluation of goodness of fit and model selection in a dynamic context. Through empirical exploration, we found that the network statistics used to fit a model are often highly correlated. For example, if there exists a triangle in a network, it is more likely to find two-stars in the network. Model identifiability should be investigated both in static ERGM models and the VCERGM to ensure appropriate model selection. For static ERGMs, one generally assesses goodness of fit through a comparison of quantitative summaries of simulated networks from the fitted model with the summaries of the observed network \citep{hunter2008goodness}. However, for dynamic networks this type of goodness of fit comparison captures only the marginal aspects of the dynamic sequence. How exactly to assess the quality of a dynamic model is still an open problem. A second avenue for future work involves adapting the varying-coefficient framework introduced here to networks with weighted edges. To do this, one can extend the exponential models of networks for integer-valued weights from \cite{krivitsky2012exponential} or to the models of networks for continuous-valued weights considered in \cite{desmarais2012statistical, wilson2017stochastic}. 

In many dynamic networks, it is often of interest to identify change-points in the network, namely points in time where the network undergoes significant local or global structural change \citep{woodall2016overview, bindu2016mining}. It would be interesting to further analyze how to utilize dynamic network models like the VCERGM to identify such changes. The test for heterogeneity that we use in the paper may provide some idea of how to formally test for a change - through identification of a change in network parameter - however, in future research we plan to purse this idea further. 

Finally, we will investigate adapting a varying-coefficient model for networks with a Markov dependency. In this paper, we investigated the dynamics of coefficients for marginal network statistics. However, this model can readily be extended to networks with a Markov dependency like that described by the first order TERGM. In general, for any non-negative integer $q$, one can model higher order dependencies between $q$ successive networks. For example if we assume the one-step Markov dependence in (\ref{eq:MarkovAssumption}), we can model the one step transition between $X_{t-1}$ and $X_{t}$ using a suite of statistics $\mathbf{g}_1(x_t, x_{t-1})$ as
\begin{equation} \label{eq:VCERGMlag}
	\begin{aligned}
\mathbb{P}(X_t = x_t \, | \, \boldsymbol{\phi}_1(t), x_{t-1}) & = \frac{\exp \{ \boldsymbol{\phi}_1(t) ^T \, \mathbf{g}_1(x_t, x_{t-1})\}}{\sum_{z \in \{0,1\}^{n_t \times n_t}} \exp\{\boldsymbol{\phi}_1(t)^T ~ \mathbf{g}_1(z, x_{t-1})\}}, \text{\hskip 1pc $x_t \in \{0,1\}^{n_t \times n_t}$}.\\
%\theta_j(t) &= \sum_{\ell = 1}^q \Phi_{j\ell} B_{\ell}(t), \text{\hskip 1pc $j = 1, \ldots, p$}\\
\end{aligned}
\end{equation}
In model (\ref{eq:VCERGMlag}), $\boldsymbol{\phi}_1(t) = \{\phi_{1k}(t), k = 1, \ldots, p\}$ can be modeled as smooth functions that describe the impact of the one-step transition statistics from $x_{t-1}$ to $x_t$. Therefore, model (\ref{eq:VCERGMlag}) effectively captures the rate of change of the temporal potential between sequential graphs rather than the rate of change of the marginal features as done in this work. Like the TERGM, we can generalize the VCERGM to a higher order Markov dependency, say order $q > 1$, by specifying appropriate transition statistics $\mathbf{g}_q(x_t, x_{t-1}, \ldots, x_{t-q})$. 

\section*{Acknowledgements}
The fMRI data were provided [in part] by the Human Connectome Project, WU-Minn Consortium (Principal Investigators: David Van Essen and Kamil Ugurbil; 1U54MH091657) funded by the 16 NIH Institutes and Centers that support the NIH Blueprint for Neuroscience Research; and by the McDonnell Center for Systems Neuroscience at Washington University. 
%Gen Li's research was partially supported by the Calderone Junior Faculty Award from the Mailman School of Public Health at Columbia University.

%\section*{Appendix}

%\addcontentsline{toc}{section}{Appendices}
%\renewcommand{\thesubsection}{\Alph{subsection}}

%\newpage
{\singlespacing
\bibliographystyle{plainnat}
% \bibliography{VCERGM_bib2.bib}

}

\bigskip
\newpage
\begin{center}
{\large\bf SUPPLEMENTARY MATERIAL}
\end{center}

\begin{description}

%%%%%%%%%%%%
% Section: Motivation %
%%%%%%%%%%%%

\item[Stochastic Equivalence under the Difference Statistic Specification:]
% Relationship between TERGM and ERGM -- Define difference statistics
Comparing the first-order TERGM with model (\ref{eq:ERGM}), we see that the TERGM is closely related to the ERGM in that it characterizes the conditional distribution of $X_t$ given $X_{t-1}$ using an ERGM representation. Perhaps not surprisingly, these two models are much more closely related than this relationship. 

Consider a simple univariate time series represented by the stochastic process $\mathbf{Z} = \{Z_1, \ldots, Z_T\}$ for $Z_t \in \mathbb{R}$. Without any other information about $\mathbf{Z}$, a natural non-parametric manner to investigate the rate of change in $\mathbf{Z}$ involves analyzing the difference between sequential observations $Z_{t-1}$ and $Z_t$, namely analyzing $\Delta(Z_t) = Z_t - Z_{t-1}$.
The analysis of $\Delta(Z_t)$ in univariate and multivariate time series is known as \emph{differencing}, and is a common first step in the analysis of time series data \citep{brockwell2013time}. In the context of the TERGM, differencing corresponds to the analysis of \emph{difference statistics}, where one specifies transition statistics of the form
\begin{equation}\label{eq:difference} \boldsymbol{g}(x_t, x_{t-1}) = \mathbf{h}(x_t) - \mathbf{h}(x_{t-1}), \hskip 1pc \text{t = 2, \ldots, T}, \end{equation}
\noindent where $\mathbf{h}: \{0,1\}^{n \times n} \rightarrow \mathbb{R}^p$ is a topological summary of an input network with $n$ vertices. Statistics of the form in (\ref{eq:difference}) can capture, for example, the differences in the edge weight of the network from time $t-1$ to $t$, or the difference in the number of triangles from one network to the next. Although incorporating {difference statistics} in the TERGM is a natural first-step in the analysis of temporal networks, it turns out that doing so is equivalent to modeling each network $X_t \in \boldsymbol{X}$ as an independent realization from the \emph{same} exponential family probability mass function. This is made precise in the next proposition.

% Proposition
\begin{proposition}\label{thm:equiv}
	Let $\boldsymbol{\mathcal{X}}$ denote the family of unweighted dynamic graph sequences on $n$ vertices with $T \geq 1$ discrete observations. Suppose that $\boldsymbol{X} = \{X_1, \ldots, X_T\} \in \boldsymbol{\mathcal{X}}$ is generated under the TERGM in (\ref{eq:tergm}), where for $t = 2, \ldots, T$
	$$X_t \mid \boldsymbol{X}_{t}^{-} \sim \mathbb{P}(X_t = x_t \mid X_{t-1} = x_{t-1}; \boldsymbol{\phi}) = \dfrac{\exp\{\boldsymbol{\phi}^T ~ \mathbf{g}(x_t, x_{t-1})\}}{\displaystyle\sum_{z \in \{0,1\}^{n \times n}} \exp\{\boldsymbol{\phi}^T ~ \mathbf{g}(z, x_{t-1})\}}.$$
	Suppose $\mathbf{g}(\cdot, \cdot) \in \mathbb{R}^p$ is a difference statistic of the form (\ref{eq:difference}) where $\mathbf{g}(x,y) = \mathbf{h}(x) - \mathbf{h}(y)$ for some $\mathbf{h}(\cdot) \in \mathbb{R}^p$. Then for all $t \geq 2$, $X_t$ is independent of $\boldsymbol{X}_{t}^{-}$ and can be generated as an independent realization of an ERGM with the following probability mass function

$$X_t \mid \boldsymbol{X}_{t}^{-} \sim \mathbb{P}(X_t = x \mid \boldsymbol{\phi}) = \frac{\exp\{\boldsymbol{\phi}^T \mathbf{h}(x)\}}{\displaystyle\sum_{z \in \{0,1\}^{n \times n}} \exp\{\boldsymbol{\phi}^T \mathbf{h}(z)\}}$$
	
\end{proposition}

% Limitation of TERGM: Cannot capture temporal heterogeneity
Proposition \ref{thm:equiv} reveals that under the difference statistic model specification, a dynamic network under the TERGM reduces to an {independent and identically distributed} sequence of graphs under a corresponding ERGM. Hence under this family of specifications, the TERGM does \emph{not} capture temporal dependence in the underlying dynamic network sequence. Although in practice one may utilize statistics that are not of the form (4), this relatively simple example motivates further investigation between the relationship of the ERGM and the TERGM.

\item[Iterative Reweighted Least Squares (IRLS):] 

% \subsection{Iterative Reweighted Least Squares (IRLS)} 

The penalized logistic regression problem for fitting a VCERGM is to maximize the following penalized log likelihood function:
\begin{equation}
\mathbf{y}^T\mathbf{H}\text{vec}(\boldsymbol{\Phi}) -\log[1+\exp\{\mathbf{H}\text{vec}(\boldsymbol{\Phi})\}] - \lambda \mathcal{P}(\boldsymbol{\Phi}). \nonumber
\end{equation}

\noindent The tuning parameter $\lambda$ controls the amount of roughness. We implement the iteratively reweighted least squares (IRLS) to fit the logistic regression with the penalty term. Consider a link function $g(\mu)= \log({\mu}/{(1-\mu)})$ and a convex function $b(\eta) = \log(1+e^{\eta})$. The IRLS without penalty term updates $\Phi$ at the $(u+1)$th iteration
\begin{equation} \label{eq:IRLS}
\text{vec}(\boldsymbol{\Phi}^{(u+1)})=(\mathbf{H}^T\mathbf{W}^{(u)}\mathbf{H})^{-1}\mathbf{H}^T\mathbf{W}^{(u)}\left\{ \mathbf{H}\text{vec}(\boldsymbol{\Phi}^{(u)})+(\mathbf{y}-\boldsymbol{\mu}^{(u)}) \cdot g'(\boldsymbol{\mu}^{(u)})\right\},
\end{equation}
where $\boldsymbol{\mu}^{(u)}=b'(\mathbf{H} \text{vec}(\boldsymbol{\Phi}^{(u)}))$ and $\mathbf{W}^{(u)}$ is a diagonal matrix with
\begin{equation}
\mathbf{W}_{(i, i)}^{(u)}={1\over b''(\mathbf{H}_{(i)}^T\text{vec}(\boldsymbol{\Phi}^{(u)}))}{1\over \{{g'}(\boldsymbol{\mu}_i^{(u)})\}^2}, \qquad i=1, 2, \ldots, (p \times q). \nonumber
\end{equation}

With the penalty term $\mathcal{P}(\boldsymbol{\Phi})$, we only need to replace $\mathbf{H}^T \mathbf{W}^{(u)} \mathbf{H}$ by $\mathbf{H}^T\mathbf{W}^{(u)} \mathbf{H} + \lambda \, (\boldsymbol{\Omega} \otimes \mathbf{I}_p)$ in (\ref{eq:IRLS}). The generalized cross validation (GCV) is used to choose the tuning parameter $\lambda$ \citep{golub1979generalized}. Namely, the $\lambda$ is a minimizer of $G(\lambda)$, which is defined as
$$
G(\lambda) = \left. \frac{1}{N} || \mathbf{y} - \mathbf{H} (\mathbf{H}^T \mathbf{H} + N \lambda \boldsymbol{\Omega})^{-1} \mathbf{H}^T \mathbf{y}) ||^2 \middle/ \Bigl\{ \frac{1}{N} \text{tr}(I -  \mathbf{H} (\mathbf{H}^T \mathbf{H} + N \lambda \boldsymbol{\Omega})^{-1} \mathbf{H}^T) \Bigr\}^2 \right. ,
$$
where $N$ is the number of rows in matrix $\mathbf{H}$.

\item[Additional Simulation Results:]

%\subsection{Additional Simulation Results}  \label{appendix:result}

Tables below show the mean and standard deviation of IAE associated with fitting ERGMs and VCERGMs to temporal networks of size 50 and 100 with 0, 1, 5, and 10 randomly missing networks. The results are from the settings (i) sinusoidal curves with $(a, b, c, d) = (1, 20, 5, 1)$ (edges) and $(a, b, c, d) = (0.6, 20, 3, 0.4)$ (reciprocity); (ii) quadratic curves with $(a, b) = (1/25^2, 0)$ (edges) and $(a, b) = (-1/30^2, 0.5)$ (reciprocity); (iii) \erdos with $p_{edges} = 0.85$; (iv) a sequence of random numbers from $N(0, 1)$ (edges) and $N(1.5, 0.6)$ (reciprocity). 

%\begin{figure}[htbp]
%\centering
%\includegraphics[width = 0.9\textwidth, height=0.35\textheight]{Figure/Simulation/Plot_n=30_nmiss=5.png}
%\caption{{\bf Parameter estimates with 30 nodes and 5 missing networks}: Estimated parameters for edges (top) and reciprocity (bottom). Black line is the true $\boldsymbol{\phi}(t)$. Red (ERGM) is for cross-sectional ERGMs, green (ERGM2) is for ad hoc  2-step procedure, and blue (VCERGM) is for VCERGM. For each method, solid line indicates the average of 100 estimated curves and the shaded band illustrates the first and third quantiles.}
%\label{figure:Simulation3}
%\end{figure}

\begin{table}[htbp]
\caption{\label{table:Simulation50} {\bf Simulation results with 50 nodes and (0, 1, 5, 10) missing networks}: Mean and standard deviation of the integrated absolute errors (IAE) for each method.}      
\centering
\resizebox{\columnwidth}{!}{
%\fbox{%
\begin{tabular}{lr|rrr|rrr}
\toprule
& \multirow{2}{*}{Missing} & \multicolumn{3}{c|}{Edges} & \multicolumn{3}{c}{Reciprocity} \\
\cline{3-8}
& & \multicolumn{1}{c}{ERGM} & \multicolumn{1}{c}{ERGM2} & \multicolumn{1}{c|}{VCERGM} & \multicolumn{1}{c}{ERGM} & \multicolumn{1}{c}{ERGM2} & \multicolumn{1}{c}{VCERGM} \\
\hline
{\bf Sinusoidal} & 0 & 10.56 (16.1) & 7.74 (16.88) & {\bf 7.61} (16.92) & 8.23 (3.22) & {\bf 4.87} (4) & 4.78 (3.96) \\
% & 1 & 10.44 (15.87) & {\bf 7.91} (16.81) & 8.28 (16.71) & 8.14 (3.22) & 5.32 (3.9) & {\bf 5.01} (3.9) \\
% & 5 & 9.65 (14.48) & 9 (16.55) & {\bf 8.63} (16.65) & 7.58 (3.1) & 6.36 (3.66) & {\bf 6.3} (3.54) \\
% & 10 & {\bf 8.34} (12.64) & 13.17 (15.62) & 13.9 (15.4) & {\bf 6.54} (2.82) & 9.26 (3.05) & 8.92 (2.95) \\
 & 1 &  & {\bf 7.91} (16.81) & 8.28 (16.71) &  & 5.32 (3.9) & {\bf 5.01} (3.9) \\
 & 5 &  & 9 (16.55) & {\bf 8.63} (16.65) &  & 6.36 (3.66) & {\bf 6.3} (3.54) \\
 & 10 &  & {\bf 13.17} (15.62) & 13.9 (15.4) &  & 9.26 (3.05) & {\bf 8.92} (2.95) \\
\hline
{\bf Quadratic} & 0 & 6.02 (11.37) & {\bf 4.14} (11.82) & 4.17 (11.8) & 5.18 (1.51) & {\bf 2.14} (1.96) & 2.42 (1.84) \\
% & 1 & 5.92 (11.18) & {\bf 4.17} (11.79) & 4.19 (11.78) & 5.08 (1.47) & {\bf 2.17} (1.97) & 2.44 (1.86)  \\
% & 5 & 5.44 (10.32) & {\bf 4.39} (11.72) & {\bf 4.39} (11.71) & 4.66 (1.35) & {\bf 2.29} (1.95) & 2.58 (1.83) \\
% & 10 & {\bf 4.86} (9.13) & 5.09 (11.54) & 5.12 (11.54) & 4.17 (1.26) & {\bf 2.77} (1.93) & 2.93 (1.84) \\
  & 1 &  & {\bf 4.17} (11.79) & 4.19 (11.78) &  & {\bf 2.17} (1.97) & 2.44 (1.86)  \\
 & 5 &  & {\bf 4.39} (11.72) & {\bf 4.39} (11.71) &  & {\bf 2.29} (1.95) & 2.58 (1.83) \\
 & 10 &  & {\bf 5.09} (11.54) & 5.12 (11.54) &  & {\bf 2.77} (1.93) & 2.93 (1.84) \\
 \hline
{\bf \erdos} & 0 & 14.7 (22.94) & {\bf 10.03} (24.32) & 10.3 (24.25) & 8.88 (1.55) & {\bf 2.77} (1.21) & 3.25 (0.89) \\
% & 1 & 14.41 (22.47) & {\bf 10} (24.33) & 10.33 (24.24) & 8.73 (1.52) & {\bf 2.79} (1.21) & 3.3 (0.88) \\
% & 5 & 13.23 (20.66) & {\bf 10.12} (24.33) & 10.46 (24.24) & 7.92 (1.38) & {\bf 2.8} (1.26) & 3.34 (0.91) \\
% & 10 & 11.72 (18.37) & {\bf 10.01} (24.34) & 10.43 (24.22) & 7.01 (1.24) & {\bf 2.97} (1.44) & 3.4 (0.85) \\
 & 1 &  & {\bf 10} (24.33) & 10.33 (24.24) &  & {\bf 2.79} (1.21) & 3.3 (0.88) \\
 & 5 &  & {\bf 10.12} (24.33) & 10.46 (24.24) &  & {\bf 2.8} (1.26) & 3.34 (0.91) \\
 & 10 &  & {\bf 10.01} (24.34) & 10.43 (24.22) &  & {\bf 2.97} (1.44) & 3.4 (0.85) \\
 \hline
{\bf Non-smooth} & 0 & 25.88 (7.26) & {\bf 25.35} (7.34) & 32.74 (6.16) & 19.87 (8.52) & 1{\bf 9.48} (8.61) & 23.25 (7.21) \\
% & 1 & {\bf 25.44} (7.17) & 26.34 (7.17) & 33.1 (6.07) & {\bf 19.72} (8.36) & 19.8 (8.58) & 23.47 (7.22) \\
% & 5 & {\bf 21.93} (6.89) & 26.46 (7.13) & 32.25 (6.2) & {\bf 17.42} (8.06) & 19.77 (8.68) & 23.61 (7.29) \\
% & 10 & {\bf 19.38} (5.39) & 34.87 (5.9) & 36.28 (5.59) & {\bf 15.27} (6.77) & 22.64 (7.81) & 23.98 (7.18) \\
 & 1 &  & {\bf 26.34} (7.17) & 33.1 (6.07) &  & {\bf 19.8} (8.58) & 23.47 (7.22) \\
 & 5 & & {\bf 26.46} (7.13) & 32.25 (6.2) &  & {\bf 19.77} (8.68) & 23.61 (7.29) \\
 & 10 &  & {\bf 34.87} (5.9) & 36.28 (5.59) &  & {\bf 22.64} (7.81) & 23.98 (7.18) \\
\bottomrule
\end{tabular}}
%}
\end{table}

\begin{table}[htbp]
	\captionsetup{width=.6\linewidth}
\caption{\label{table:Simulation100} {\bf Simulation results with 100 nodes and (0, 1, 5, 10) missing networks}: Mean and standard deviation of the integrated absolute errors (IAE) for each method.}     
\centering
\resizebox{\columnwidth}{!}{
%\fbox{%
\begin{tabular}{lr|rrr|rrr}
\toprule
& \multirow{2}{*}{Missing} & \multicolumn{3}{c|}{Edges} & \multicolumn{3}{c}{Reciprocity} \\
\cline{3-8}
& & \multicolumn{1}{c}{ERGM} & \multicolumn{1}{c}{ERGM2} & \multicolumn{1}{c|}{VCERGM} & \multicolumn{1}{c}{ERGM} & \multicolumn{1}{c}{ERGM2} & \multicolumn{1}{c}{VCERGM} \\
\hline
{\bf Sinusoidal} & 0 & 25.96 (35.76) & 25.54 (36.02) & {\bf 25.51} (36.04) & 8.88 (6.62) & 8.19 (7.03) & {\bf 8.14} (7.01) \\
% & 1 & 25.66 (35.17) & {\bf 25.38} (36.09) & 25.44 (36.05) & 8.83 (6.59) & 8.45 (6.82) & {\bf 8.39} (6.85) \\
% & 5 & {\bf 23.54} (32.2) & 26.3 (35.63) & 26.28 (35.63) & {\bf 8.31} (6.28) & 8.86 (6.61) & 8.78 (6.5) \\
% & 10 & {\bf 20.38} (28.18) & 29.13 (34.25) & 29.29 (34.16) & {\bf 7.28} (5.7) & 11.69 (4.91) & 11.61 (4.85) \\
 & 1 &  & {\bf 25.38} (36.09) & 25.44 (36.05) &  & 8.45 (6.82) & {\bf 8.39} (6.85) \\
 & 5 &  & 26.3 (35.63) & {\bf 26.28} (35.63) &  & 8.86 (6.61) & {\bf 8.78} (6.5) \\
 & 10 &  & {\bf 29.13} (34.25) & 29.29 (34.16) &  & 11.69 (4.91) & {\bf 11.61} (4.85) \\
\hline
{\bf Quadratic} &  0 & 15.45 (27.98) & {\bf 14.96} (28.23) & 15.04 (28.18) & 4.29 (3.36) & {\bf 3.13} (3.78) & 3.24 (3.63) \\
% & 1 & 15.2 (27.51) & {\bf 14.87} (28.24) & 14.94 (28.21) & 4.17 (3.21) & {\bf 3.23} (3.71) & 3.32 (3.58) \\
% & 5 & {\bf 14.04} (25.3) & 15 (28.13) & 15.03 (28.11) & 3.84 (2.99) & {\bf 3.34} (3.71) & 3.41 (3.6) \\
% & 10 & {\bf 12.43} (22.45) & 15.63 (27.81) & 15.63 (27.81) & {\bf 3.41} (2.75) & 3.77 (3.66) & 3.83 (3.55) \\
 & 1 &  & {\bf 14.87} (28.24) & 14.94 (28.21) &  & {\bf 3.23} (3.71) & 3.32 (3.58) \\
 & 5 &  & {\bf 15} (28.13) & 15.03 (28.11) &  & {\bf 3.34} (3.71) & 3.41 (3.6) \\
 & 10 &  & {\bf 15.63} (27.81) & {\bf 15.63} (27.81) &  & {\bf 3.77} (3.66) & 3.83 (3.55) \\
 \hline
{\bf \erdos} & 0 & 37.23 (47.52) & {\bf 36.48} (48.06) & 36.61 (47.97) & 3.6 (0.92) & {\bf 1.02} (0.5) & 1.48 (0.39) \\
% & 1 & 36.48 (46.58) & {\bf 36.46} (48.07) & 36.59 (47.99) & 3.54 (0.91) & {\bf 0.99} (0.46) & 1.5 (0.4) \\
% & 5 & {\bf 33.59} (42.73) & 36.67 (47.94) & 36.78 (47.87) & 3.21 (0.8) & {\bf 1} (0.47) & 1.5 (0.39) \\
% & 10 & {\bf 29.81} (37.94) & 36.42 (48.02) & 36.58 (47.91) & 2.94 (0.82) & {\bf 1.27} (0.56) & 1.67 (0.45) \\
 & 1 &  & {\bf 36.46} (48.07) & 36.59 (47.99) &  & {\bf 0.99} (0.46) & 1.5 (0.4) \\
 & 5 &  & {\bf 36.67} (47.94) & 36.78 (47.87) &  & {\bf 1} (0.47) & 1.5 (0.39) \\
 & 10 &  & {\bf 36.42} (48.02) & 36.58 (47.91) &  & {\bf 1.27} (0.56) & 1.67 (0.45) \\
 \hline
{\bf Non-smooth} & 0 & {\bf 32.48} (19.59) & 32.64 (19.57) & 37.05 (18.12) & {\bf 27.37} (15.95) & 27.51 (15.8) & 28 (14.77) \\
% & 1 & {\bf 31.95} (19.29) & 33.48 (19.26) & 37.57 (17.9) &  {\bf 27.16} (15.64) & 27.78 (15.74) & 28.31 (14.77) \\
% & 5 & {\bf 28.12} (18.32) & 33.25 (19.35) & 36.69 (18.2) & {\bf 24.58} (14.82) & 27.61 (15.77) & 28.37 (14.74) \\
% & 10 & {\bf 24.02} (15.03) & 40.36 (17.2) & 41.84 (16.62) & {\bf 20.96} (12.7) & 29.54 (14.52) & 29.37 (14.18) \\
 & 1 &  & {\bf 33.48} (19.26) & 37.57 (17.9) &   & {\bf 27.78} (15.74) & 28.31 (14.77) \\
 & 5 &  & {\bf 33.25} (19.35) & 36.69 (18.2) &  & {\bf 27.61} (15.77) & 28.37 (14.74) \\
 & 10 &  & {\bf 40.36} (17.2) & 41.84 (16.62) &  & 29.54 (14.52) & {\bf 29.37} (14.18) \\
\bottomrule
\end{tabular}}
%}
\end{table}

\item[R-package for  VCERGM:] R-package VCERGM containing code to fit the VCERGM, conduct hypothesis test for heterogeneity, and compare the VCERGM with cross-sectional ERGMs as described in the article. The package also contains political voting network dataset as provided as an example in Section \ref{subsec:political} in the article.

\end{description}

\end{document}